\documentclass[3p,twocolumn]{elsarticle}
\usepackage{amsmath}
\usepackage{amssymb}
\usepackage{graphicx}
\usepackage{dcolumn}
\usepackage{bm,bbm}
\usepackage{color}
\usepackage{placeins}
\usepackage{multirow}
\usepackage{hyperref}
\hypersetup{
    colorlinks,
    citecolor=black,
    filecolor=black,
    linkcolor=black,
    urlcolor=black
}

\begin{document}

\title{Generation of 10-m-lengthscale plasma columns by resonant and off-resonant laser pulses}%

\author[1]{G. Demeter\corref{cor}}
\author[2]{J. T. Moody\fnref{first}}
\author[1]{M. \'{A}. Kedves}
\author[2,3]{F. Batsch}
\author[2]{M. Bergamaschi}
\author[3]{V. Fedosseev}
\author[3]{E. Granados}
\author[2]{P. Muggli}
\author[3]{H. Panuganti}
\author[2,3]{G. Zevi Della Porta}

\address[1]{Wigner Research Centre for Physics, Budapest, Hungary}
\address[2]{Max Planck Institute for Physics, Munich, Germany}
\address[3]{CERN, Geneva, Switzerland}

\cortext[cor]{Corresponding author, \texttt{demeter.gabor@wigner.hu}}
\fntext[first]{Now at: Lawrence Livermore National Laboratory, California, USA}

\date{\today}

\begin{abstract}
Creating extended, highly homogeneous plasma columns like that required by 
plasma wakefield accelerators can be a challenge.
We study the propagation of ultra-short, terawatt power ionizing laser pulses in a 10-meter-long rubidium vapor and the plasma columns they create.  We perform experiments and numerical simulations for pulses with 780 nm central wavelength, which is resonant with the D$_2$ transition from the ground state of rubidium atoms, as well as for pulses with 810 nm central wavelength, some distance from resonances. We measure transmitted energy and transverse width of the pulse and use schlieren imaging to probe the plasma column in the vapor close to the end of the vapor source.
We find, that resonant pulses are more confined in a transverse direction by the interaction than off-resonant pulses are and that the plasma columns they create are  more sharply bounded.
Off-resonant pulses leave a wider layer of partially ionized atoms and thus lose more energy per unit propagation distance. Using experimental data, we estimate the energy required 
to generate a 20-meter-long plasma column and conclude 
that resonant pulses are much more suitable for creating a long, homogeneous plasma. 
\end{abstract}

\maketitle

\section{Introduction}
\label{intro}

The propagation of high-power, ionizing laser pulses in gases has been studied extensively for decades, under a wide range of conditions \cite{Couairon2007, Berge2007, Kandidov2009}. Phenomena such as self-focusing, filamentation or super-continuum generation were investigated and numerous applications for these phenomena devised (remote sensing, nonlinear spectroscopy,  lightning protection, etc.). One particular application is the creation of plasma columns to be used in wakefield particle acceleration. Plasma wakefield accelerators are capable of accelerating electrons (or positrons) in the strong electric fields sustained by plasma waves, \cite{Joshi2003, Leemans2009, Malka2008, Esarey2009, Tajima2020}. Since the accelerating gradients can be up to three orders of magnitude larger than in conventional particle accelerators, wakefield acceleration may be a replacement for established accelerator technology in compact, cost-effective particle accelerators for science and commercial applications. A multitude of scientific and technological challenges need to be tackled (only one of which is the creation of the necessary plasma column), but wakefield accelerators and their applications are advanced constantly by numerous research groups worldwide \cite{Albert2016,Hidding2019,Adli2019v3,Adli2022}.
One notable example of such a wakefield accelerator project is the Advanced Wakefield Experiment (AWAKE) at CERN (European Laboratory for Particle Physics in Geneva, Switzerland), the first proton driven wakefield accelerator \cite{Gschwendtner2016,Adli2018v2}. 

At the heart of the AWAKE device is a 10-meter-long plasma column with finely engineered plasma density, which is essential for accelerator operation. High energy proton bunches from the Super Proton Synchrotron facility at CERN interact with this plasma to create large amplitude
wakefields, which in turn can accelerate electron bunches well into the GeV domain.
This plasma column is created by starting from a vapor of rubidium atoms with precisely tuned density distribution along the vapor source axis and propagating a terawatt (TW) power laser pulse along the vapor to achieve exactly one-electron ionization of atoms with a probability very close to one. Precisely tuned vapor density thus yields a finely tuned plasma density. Achieving complete one-electron ionization is facilitated by the fact that the 780 nm wavelength laser pulse is resonant with the $\mathrm{5s^2S_{1/2}}\rightarrow\mathrm{5p^2P_{3/2}}$ transition of the valence electron from the atomic ground state (the D$_2$ line) and then further from $\mathrm{5p^2P_{3/2}}$ to $\mathrm{5d^2D_{5/2}}$, $\mathrm{5d^2D_{3/2}}$  states.     
These single-photon resonances also have a major impact on the propagation of the ionizing laser pulse -- a question studied only very recently in the context of ultra-short pulse propagation \cite{Demeter2019}. 
It has been suggested \cite{Demeter2021}, that it is predominantly the single-photon resonances that give rise to a strong but saturable nonlinearity 
which can be very advantageous for the propagation of the ionizing pulse. In particular, theoretical indication was that 
due to resonant self-focusing of the pulse by the vapor, pulse energy in the tail is channeled very effectively within the plasma created by the leading edge of the pulse. 
The plasma column therefore becomes longer and more sharply bounded when a 780 nm ionizing pulse is used. 

It is important to note that the laser pulse in this scheme is not intense enough to drive wakefields in the plasma (such as in laser wakefield acceleration). Field strengths are orders of magnitude smaller and therefore associated nonlinear propagation effects (such as relativistic self-focusing) are absent. The scheme is related to plasma column 
formation for laser wakefield acceleration with the help of pre-formed plasma waveguides \cite{Miao2020,Gupta2022}. However, as plasma densities are orders of magnitude smaller, inhomogeneities of plasma dispersion itself play little role during propagation.

Here we present an experimental investigation of plasma column generation by resonant and off-resonant TW laser pulses. We compare the propagation properties of pulses with 780 nm central wavelength (the rubidium D$_2$ line) and pulses with 810 nm central wavelength in the 10-meter-long vapor source at the CERN AWAKE site. We also measure the properties of the created plasma column close to the downstream end of the vapor source using schlieren imaging. Using measurement data, an extension of the plasma column creation with the same laser apparatus to 20 meter length is considered. 
We generalize the theory derived for resonant pulse propagation \cite{Demeter2019} to treat both resonant and off-resonant pulses with the same equations and we perform numerical simulations to study the pulse propagation process. Comparing measurement results with calculations we show that the predictions of our theory are qualitatively correct for both resonant and off-resonant propagation. However, some quantitative discrepancies between simulation and experiment remain in certain respects. The primary application of our results is in wakefield accelerator design, but they are also interesting for 
any application associated with the propagation of high-power laser pulses 
such as the creation of long plasma channels for lightning protection \cite{Houard2023} or remote sensing applications where resonances play an important role.

\section{Experiment}

\subsection{Laser propagation experiment apparatus}

\begin{figure}[htb]
\centering
\includegraphics[width=0.47\textwidth]{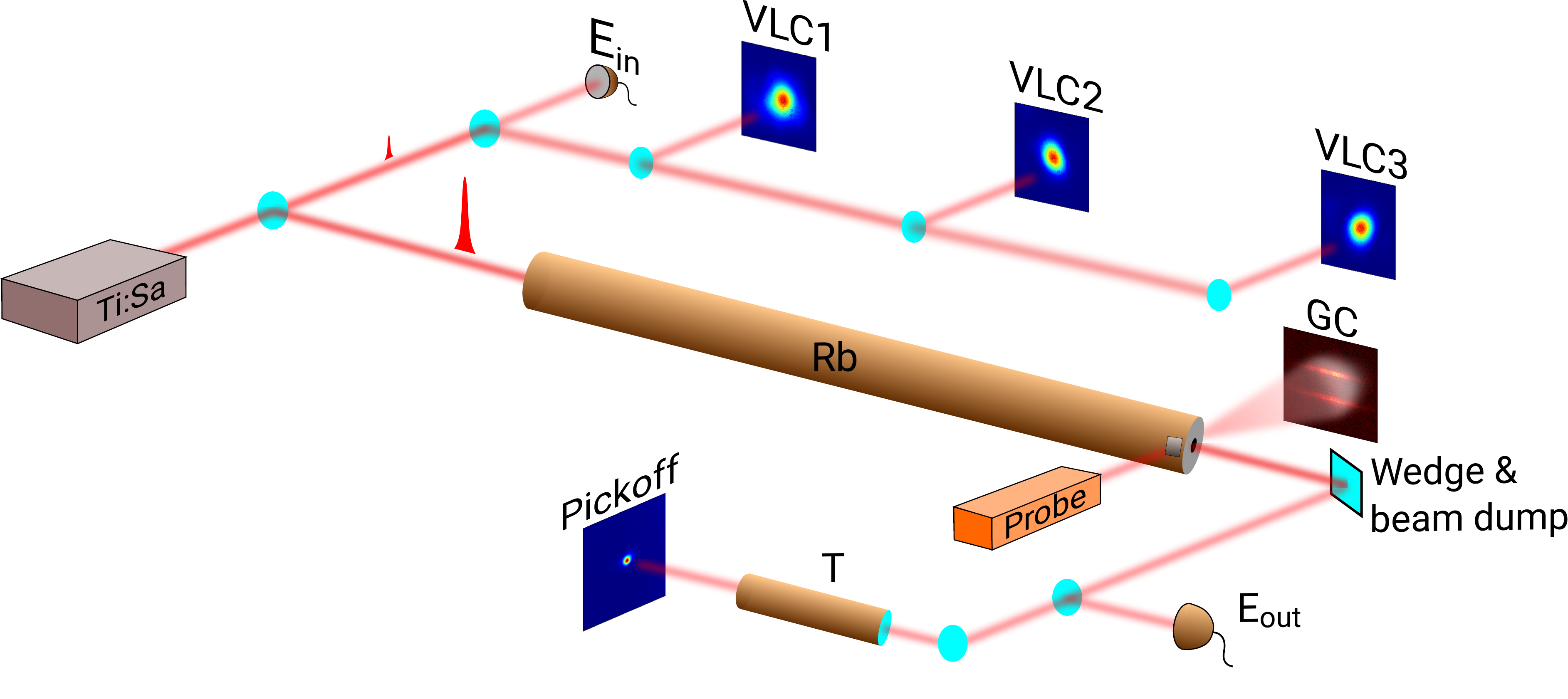}
\caption{Sketch of the experimental setup. Pulses from a TW laser system (Ti:Sa) propagate along the rubidium vapor source (Rb). About 1\% of the pulse energy is deflected to the virtual laser line to be monitored by an energy meter ($E_{in}$) and cameras (VLC1-3). About 0.5\% of the transmitted pulse is reflected off a wedge to an energy meter ($E_{out}$) and through an imaging telescope (T) to a camera (Pickoff). A transverse probe beam close to the downstream end of the vapor source is used for schlieren imaging on a gated camera (GC).}
\label{fig_setup}       
\end{figure}

\begin{figure}[htb]
\centering
\includegraphics[width=0.47\textwidth]{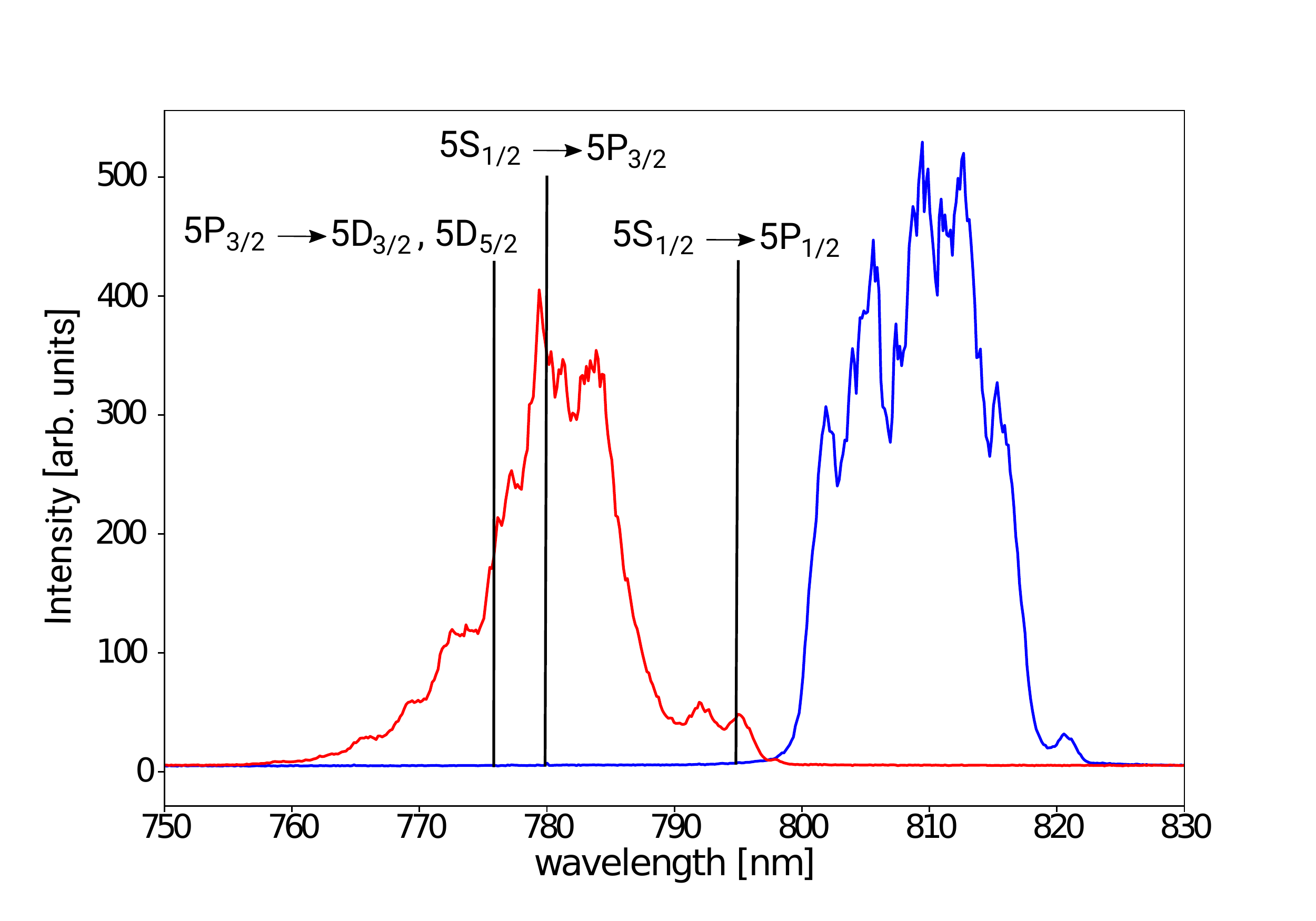}
\caption{Measured spectrum of the resonant (red) and off-resonant (blue) laser pulses. Single-photon resonances to the
$\mathrm{5p^2P_{1/2}}$ (795 nm) and $\mathrm{5p^2P_{3/2}}$ (780 nm) first excited states are marked by black lines, as well as the
resonance from $\mathrm{5p^2P_{3/2}}$ to higher lying excited states  
$\mathrm{5d^2D_{3/2}}$ and $\mathrm{5d^2D_{5/2}}$ (776 nm - see also Fig. \ref{fig_Rblevels}).}
\label{fig_spectrum}       
\end{figure}

Experiments were performed at the AWAKE site at CERN, with the rubidium vapor source of the wakefield accelerator device \cite{Gschwendtner2016, Muggli2017, Gschwendtner2022}, a 10 m long, 4 cm diameter, temperature controlled steel tube that contains the Rb vapor. A schematic drawing of the experimental setup can be seen on Fig. \ref{fig_setup}. A TW class Ti:Sa laser system  supplied $\sim$120 fs duration, $\sim$150 mJ energy pulses for the ionization of the vapor. A mismatched,  $\sim$40 m effective focal length telescope was used to focus the pulses into 
the vapor source through a 10-mm-diameter aperture. The waist diameter was $w\approx 1.8$ mm full width at half maximum (FWHM), the waist location was tuned near the center of the vapor source. 
In half of the measurements, spectral shaping methods were implemented to confine the spectrum of a wide bandwidth Ti:Sa oscillator to a region around 780 nm wavelength, precisely the wavelength of the Rb D$_2$ resonance line, the $\mathrm{5s^2S_{1/2}}\rightarrow\mathrm{5p^2P_{3/2}}$ transition from the atomic ground state (Fig. \ref{fig_spectrum}). The spectrum also had significant intensity
at the 795 nm $D_1$ resonance line ($\mathrm{5s^2S}_{1/2}\rightarrow\mathrm{5p^2P}_{1/2}$ transition) and the 776 nm transitions to higher lying excited states 
($\mathrm{5p^2P}_{3/2}\rightarrow\mathrm{5d^2D}_{3/2}$ and $\mathrm{5d^2D}_{5/2}$), similarly to our previous experiments \cite{Demeter2021}. 
In the other half of the measurements, spectral shaping was used to obtain laser pulses with the central wavelength shifted away from these resonances.
The resulting amplified pulse spectrum had a central wavelength of 810 nm and virtually no power at the Rb resonance wavelengths
(Fig. \ref{fig_spectrum}). These \textit{off-resonant} pulses  had about 20\% less maximum input energy than the 780 nm central wavelength \textit{resonant} pulses.

Precise temperature control of the vapor source reservoir and walls made it possible to create a regulated,  homogeneous, constant density Rb vapor ($\delta\rho/\rho<0.5\%$) in a $\mathcal{N}=10^{14}-10^{15}$ cm$^{-3}$ range. We measured rubidium density at the upstream end of the vapor source using white-light interferometry \cite{Oz2014,Plyushchev2017,Batsch2018}. The laser pulse energy was regulated by a half-waveplate and two thin-film polarizers at Brewster angle between the last amplifier and the compressor.  
A ``virtual'' laser line was set up using the transmission from one of the transport mirrors in the laser line upstream of the vapor source. This line had cameras that recorded the laser pulse transverse energy distribution at propagation distances corresponding to the entrance (VLC1), center (VLC2) and exit (VLC3) of the vapor source (Fig. \ref{fig_setup}). 
These images represent propagation of the laser pulse in vacuum, for comparison with its propagation through the vapor source and for monitoring the focusing.
An energy meter was also placed in the virtual line to measure pulse input energy $E_{in}$, calibrated using a direct energy meter when the vapor source vacuum system was open.

Downstream of the vapor source, a wedge before the beam dump diverted $\sim0.5$\% of the transmitted laser pulse to the output energy meter $E_{out}$ and an imaging system that created an image of the vapor source output aperture on the pickoff camera. This was used to record the ionizing pulse transverse energy profile after propagating through the vapor. We calibrated $E_{out}$ readings to $E_{in}$ values by a series of measurements performed with only residual rubidium vapor in the chamber
($\mathcal{N}\ll 0.5\cdot 10^{14}$ cm$^{-3}$), which is estimated to absorb energy from the pulse well below the $\sim$ mJ noise floor of the output energy measurement and affect negligibly the pulse energy distribution. We used the same measurements to scale the size of the pickoff images to the known size of the virtual exit camera image. Various filters were used on each of the
virtual laser line cameras and the pickoff camera to prevent image saturation. 
The supplemental material of reference \cite{Demeter2021} contains a detailed drawing of the experimental setup for the energy and transmitted pulse measurement, the laser virtual line and a description for the calibration procedure for the output energy and transmitted pulse transverse profile.

In order to examine the difference between the interaction of resonant and off-resonant laser pulses with the Rb vapor,
we performed propagation measurements for several vapor densities, switching the laser spectrum each time to take data with  both resonant and off-resonant ionizing laser pulses. Measurements were done for $\mathcal{N}\approx2\cdot10^{14}\mathrm{~cm^{-3}}$, $\mathcal{N}\approx5\cdot10^{14}\mathrm{~cm^{-3}}$,  $\mathcal{N}\approx7\cdot10^{14}\mathrm{~cm^{-3}}$ and
$\mathcal{N}\approx1\cdot10^{15}\mathrm{~cm^{-3}}$ vapor densities, the precise value differing by less than $\pm2\%$ for
the corresponding resonant pulse / off-resonant pulse measurements.
For $\mathcal{N}\approx5\cdot10^{14}\mathrm{~cm^{-3}}$ and $\mathcal{N}\approx7\cdot10^{14}\mathrm{~cm^{-3}}$ vapor densities, we also measured the properties of the plasma column close to the downstream end of the vapor source using schlieren imaging together with the transmitted pulse measurements.

\subsection{Schlieren imaging of the plasma column}

\begin{figure}[htb]
\centering
\includegraphics[width=0.47\textwidth]{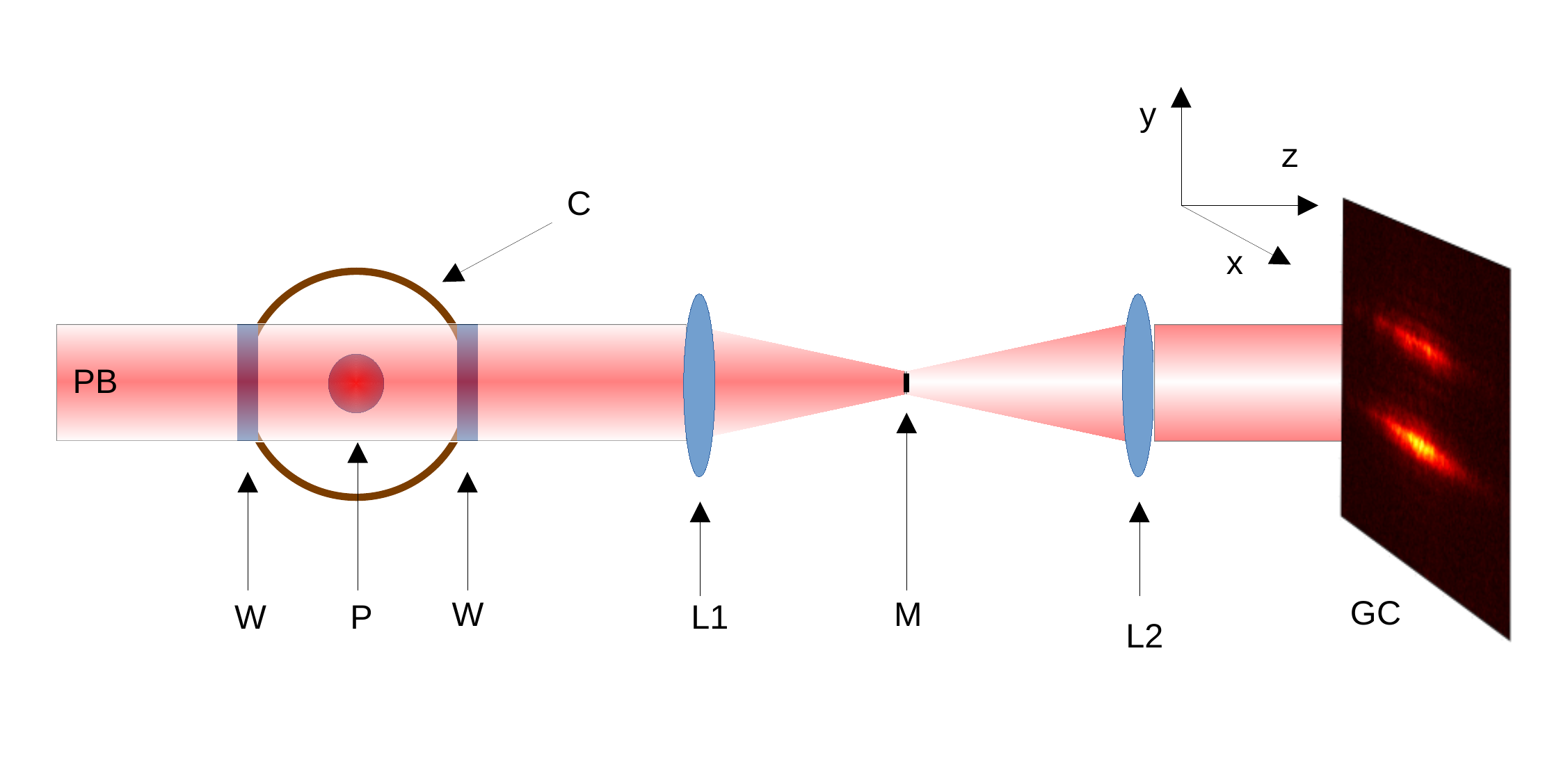}
\includegraphics[width=0.47\textwidth]{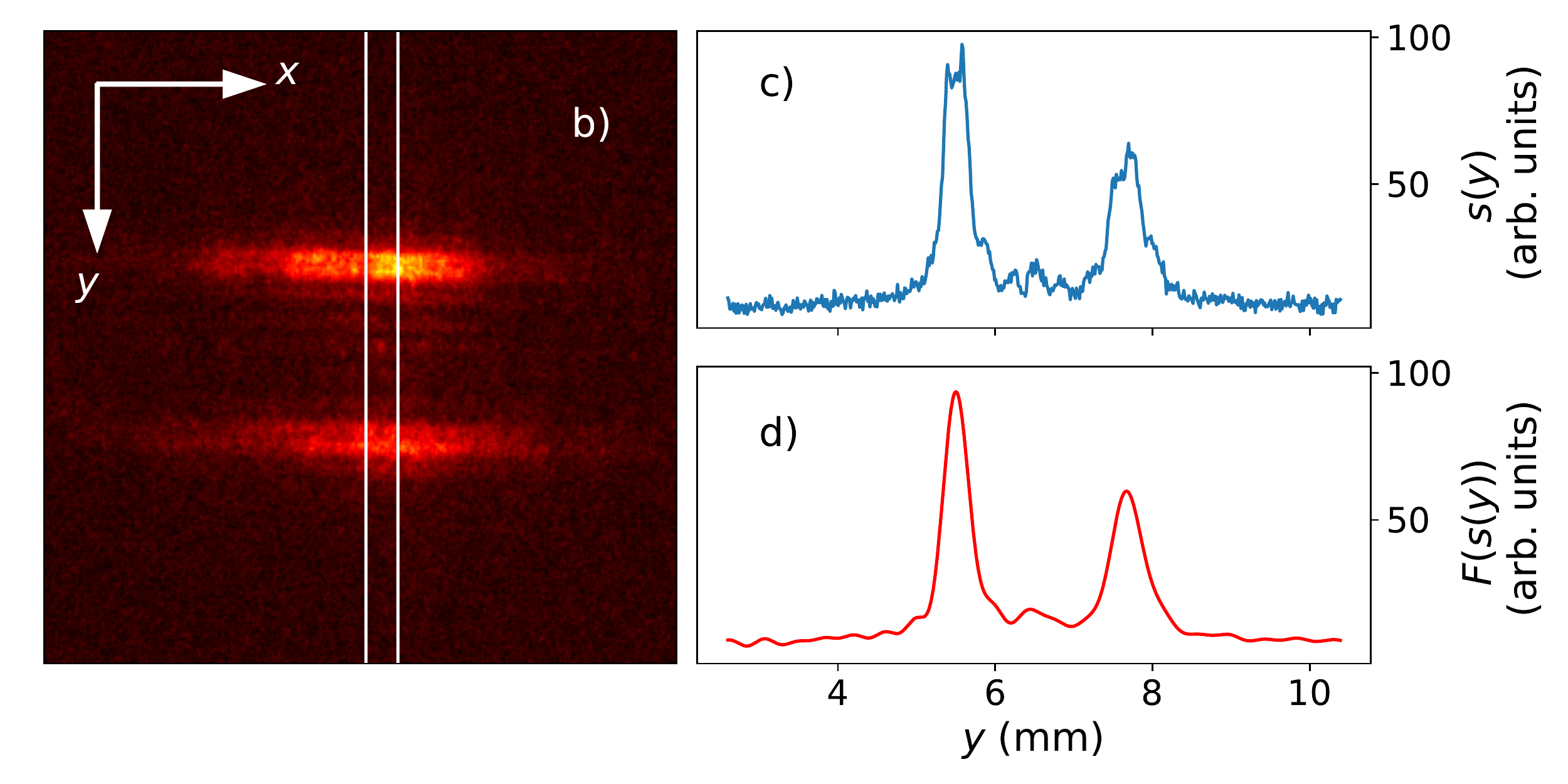}
\caption{a) Sketch of the schlieren imaging setup. A probe beam (PB) traverses the vapor source (C) cross section through a pair of sapphire view ports (W)  along the $z$ direction, sampling the refractive index of the vapor and the plasma column (P). With two lenses (L1, L2) in a $4f$ setup and a mask (M) between them, a schlieren image is created on the gated camera (GC). b) A narrow region of the gated camera image is extracted and averaged to obtain the signal c), which is then frequency filtered before being evaluated d).}
\label{fig_schlierensetup}       
\end{figure}

Schlieren imaging is 
a very sensitive method to measure refraction index changes in transparent media, used predominantly in aeronautics and fluid mechanics \cite{Schlierenbook}, but also employed regularly to investigate laser induced plasma \cite{Iwase1998,Clayton1998,Honda2000,Veloso2006,Batani2019}. Recently it was tested to probe atomic excitation in rubidium vapor \cite{Bachmann2018} and to measure plasma column properties in ionized rubidium \cite{Kerscher2021}.
In our setup, the $\lambda_p=780.311$ nm (in vacuum) continuous wave diode laser probe, tuned close to the precise value of the $\lambda_{D2}=780.241$ nm (in vacuum) D$_2$ resonance line, crossed the vapor source in a transverse direction through a pair of sapphire view ports, as seen on Fig. \ref{fig_schlierensetup}. The vapor refractive index contribution is thus
$\delta n= 10^{-4} - 10^{-3}$, due to anomalous dispersion by ground state atoms. This is large enough for detection despite the low density of vapor compared to standard atmospheric densities and it is 3-4 orders of magnitude larger than the refractive index contribution due to plasma dispersion at the same density. When the ionizing laser creates plasma, the population of the atomic ground state is reduced significantly, so the refractive index changes locally. A spatially dependent phase shift is imprinted upon the probe beam during its transit.

We placed two 75 cm focal length lenses in a $4f$ setup \cite{SalehTeich} after the vapor cell and circular mask with a diameter of $D_m=1.5$ mm at the back focal plane of the first lens. We used a gated, image-intensified camera
(Andor iStar DH334T-18F-73) to retrieve images of the probe light, triggered 100 ns after the ionizing pulse and timed to collect light for a duration of 100 ns. With this timing, atoms excited, but not ionized by the laser will return to the ground state by spontaneous decay, while plasma recombination (1-10 $\mu$s timescale at these densities) will not alter the plasma density considerably yet. Thus probing the absence of ground state atoms yields information on the plasma density.

A typical schlieren image captured by the measurement can be seen on Fig. \ref{fig_schlierensetup} b). 
Given that the properties of the plasma column are constant on the mm scale along the ionizing pulse propagation direction $x$, the $w=3.1$ mm FWHM diameter probe beam samples a $y-z$ dependent vapor density, 
the $x$ dependence on the image is due only to the probe beam intensity variation. 
Therefore it is convenient to take a rectangular region from the image, narrow in the $x$ direction  around the center of the probe beam and average along $x$ to improve the signal-to-noise ratio. The 1D $s(y)$ curve obtained is the schlieren signal that we use to analyze the plasma column (shown in Fig. \ref{fig_schlierensetup} c) ).

\subsection{Inferring plasma column properties}

The parameters of the schlieren probe beam were determined in a series of measurements with the mask removed and only residual vapor density in the vapor source. 
Given these parameters and the precise data of the anomalous dispersion of the vapor \cite{Siddons2008}, the schlieren signal $s(y)$ can also be calculated theoretically using standard formulas of Fourier optics \cite{SalehTeich} for any given plasma density distribution $N_p(y,z)$. (Note however, that at the vapor densities considered, the homogeneous lineshape function in \cite{Siddons2008} must be augmented by a collision broadening term \cite{vanLange2020},
the magnitude of which contains a constant known experimentally to much lower accuracy than the spontaneous decay rate.)
To obtain information on the extent of the plasma column from $s(y)$, we start by assuming some sensible profile for the plasma density and calculating the \textit{theoretical} schlieren signal.  According to theory \cite{Demeter2019,Demeter2021} and experiment \cite{Adli2019v2}, when the ionizing pulse is powerful enough, the ionization fraction at the center of the column is very close to one, i.e. plasma density is saturated at the initial vapor density. Thus we assume an axisymmetric plasma density of the form: 
\begin{equation}
\label{eq_plasmadensity}
 {N}_{p}(y,z) = \left\{ 
 \begin{aligned}
 &\mathcal{N} P_{max}, &\mathrm{~if~} r\leq r_0,\\
 &\mathcal{N} P_{max}\exp\left(-\frac{(r-r_0)^2}{t_0^2}\right) ,&\mathrm{~if~} r>r_0.
  \end{aligned}\right.
\end{equation}
Here $\mathcal{N}$ is the rubidium vapor density, $r$ is the distance from the column center, which is located at coordinates $(y_0,z_0=0)$ in the $y-z$ plane. (A nonzero value of $z_0$ does not change the schlieren signal because probe light phase modulation and absorption arise as dielectric parameter integrals along $z$.)  $r_0$ is the radius of the plasma column core, $P_{max}$ is the maximum ionization fraction at the center and $t_0$ is the sheath layer width parameter, a value that characterizes the width of the transition region between the core and the completely unionized vapor of neutral atoms. When the vapor is not ionized completely at the center ($P_{max}<0$), we expect $r_0 \approx 0$, and a clear maximum of ionization fraction at the center. 
A substantial value of $r_0$ is only compatible with 
$P_{max}\approx1$, the saturation of ionization in the core. The reason is that an extended region of constant ionization fraction can be realized either if we realize a sizable region in the transverse plane where the time-dependent laser pulse intensity is the same, or if the fluence is simply high enough for the ionization fraction to saturate to 1. The former is very hard to imagine in a system with a propagating pulse (self-focusing, diffraction).

The choice of the plasma profile Eq. \ref{eq_plasmadensity} (in particular the Gaussian decay function outside the core) is motivated by the fact that in the limit of less intense fields (multiphoton ionization) and pulse profile undistorted by nonlinear self-focusing (at the entrance of the vapor cell), a Gaussian beam profile will give rise to a Gaussian plasma profile.  
For the general case, (pulse profile already distorted by the nonlinear interaction during propagation) the sheath layer width parameter $t_0$ can simply be regarded as a parameter of a function fit. It will depend on the transverse fluence distribution of the propagating laser pulse and ultimately it helps characterize the strength of the self-focusing effect of the vapor. Note that this width is not related to the plasma Debye length, as the ionization takes place on the 100 fs timescale, much shorter than the timescale for any plasma dynamical phenomena. 

Calculating the schlieren signal $s(y)$ for plasma columns described by Eq. \ref{eq_plasmadensity} with a range of sensible parameter values, one can verify that the plasma gives rise to a double peaked structure in $s(y)$, similar to the experimental signal (Fig. \ref{fig_schlierensetup}). The two peaks are due to the probe light phase modulation varying in space most near the top and bottom edges of the plasma column. 
It is convenient to frequency-filter $s(y)$ numerically with a low pass filter to 
remove spatial frequencies $f_y\geq D_{mask}/(\lambda_p l)$, as the interference of light passing above and below the mask edges distorts the peaks somewhat ($l=75$ cm is the distance between the mask and the second lens). 
We can compute the locations $y_1,y_2$ of the two largest peaks of the filtered schlieren signal $\mathcal{F}\left(s(y)\right)$, as well as the peak widths $\sigma_1,\sigma_2$ and peak heights $A_1,A_2$ using (a slightly tweaked version of) the \texttt{find\_peaks} function of \texttt{SciPy} \cite{SciPy2020}. It is then possible to verify that
the core center location $y_0$ can be determined by $y_0=(y_1+y_2)/2$ with very good accuracy. Furthermore, defining the peak distance $\Delta$ and normalized peak width $W$:
\begin{eqnarray}
 \Delta & = & |y_2-y_1|\\
 W & = & \frac{A_1\sigma_1+A_2\sigma_2}{A_1 + A_2}
\end{eqnarray}
we observe that these quantities are, to a very good approximation, linear functions of $r_0$ for fixed $t_0$ and vice versa, while they do not depend on $y_0$ at all. Therefore we write their functional dependence in the following form:
\begin{eqnarray}
 \Delta &= M_{12}r_0 t_0 + M_1 r_0 + M_2 t_0 + B\nonumber\\
 W &= Q_{12} r_0 t_0 + Q_1 r_0 + Q_2 t_0 + P
 \label{eq_schlierenconsts}
\end{eqnarray}
and use a set of signals calculated with varying $r_0$, $t_0$ and $y_0$ to determine, 
using a fitting procedure, the set of constants $\{M_1, M_2, M_{12}, B, Q_1, Q_2, Q_{12}, P\}$
from $\Delta(r_0,t_0)$ and $W(r_0,t_0)$. Once known, we can evaluate any schlieren signal we obtain from the experiment by spatial frequency filtering to get rid of interference (and high-frequency noise components) and using peak finding to determine $y_0$, $\Delta$ and $W$.
We can then associate with the signal the plasma core radius $r_0$ and the sheath layer width $t_0$
obtained by inverting the relations Eqs. \ref{eq_schlierenconsts}. 
In a sense, we can regard $r_0$ and $t_0$ as the parameters of a function fit 
of the form Eq. \ref{eq_plasmadensity} on the experimental plasma distribution.

For some parameter combinations, the two-peaked structure may be absent, so we cannot associate $r_0,t_0$ values with $s(y)$ using the procedure --- in this case the image cannot be evaluated.
This may happen if sufficiently large values of $y_0$ and/or $r_0$ and/or $t_0$ combine such that one or both of the plasma edges lie far to one side where the probe beam is already too weak. This may also happen if $r_0$ and $t_0$ are both small and the two peaks are not separated. With the present measurement, $r_0+t_0 \gtrsim 0.3$ mm is required for a reliable separation of the peaks. While frequency filtering the schlieren signal to mitigate interference effects 
increases the accuracy of the evaluation, it carries a price. Very sharp plasma boundaries ($t_0\leq 0.1$ mm) give rise to narrow peaks in the schlieren image and are distorted by the filtering we use. This effectively sets the lower limit on the sheath width we can reliably evaluate. Any $t_0$ below this limit will be measured as $t_0\approx0.1$ mm.

Note that $\Delta$ and $W$ do not depend on the overall magnitude of the signal. This is convenient because vapor absorption is not known to a high accuracy due to collision broadening and also because probe laser power was not monitored continuously. The evaluation process just described yields no information on $P_{max}$ -- but when there is a measurable plasma column core ($r_0>0$) we can safely assume $P_{max}=1$. We further note that the fit coefficients (with given beam parameters) depend somewhat on the vapor density, so for the evaluation of any measurement, the corresponding set of theoretical samples must be computed and the fitting parameters determined.

Estimating the plasma parameters using machine learning methods has also been tested previously \cite{Biro2023}. Deep neural networks were trained using a large number of calculated signals to recover the underlying parameters and they proved more accurate than the fitting procedure in this paper for the calculated signals. However, for the actual experimental data, their predictions exhibited large fluctuations at times, most probably because real plasma columns are not axially symmetric and may have slightly different sheath thickness at the top and the bottom edge. The present evaluation method proves more robust with respect to this circumstance.

\section{Resonant vs. off-resonant pulse propagation}

\subsection{Experimental observations}

\begin{figure}[htb]
\centering
\includegraphics[width=1\columnwidth]{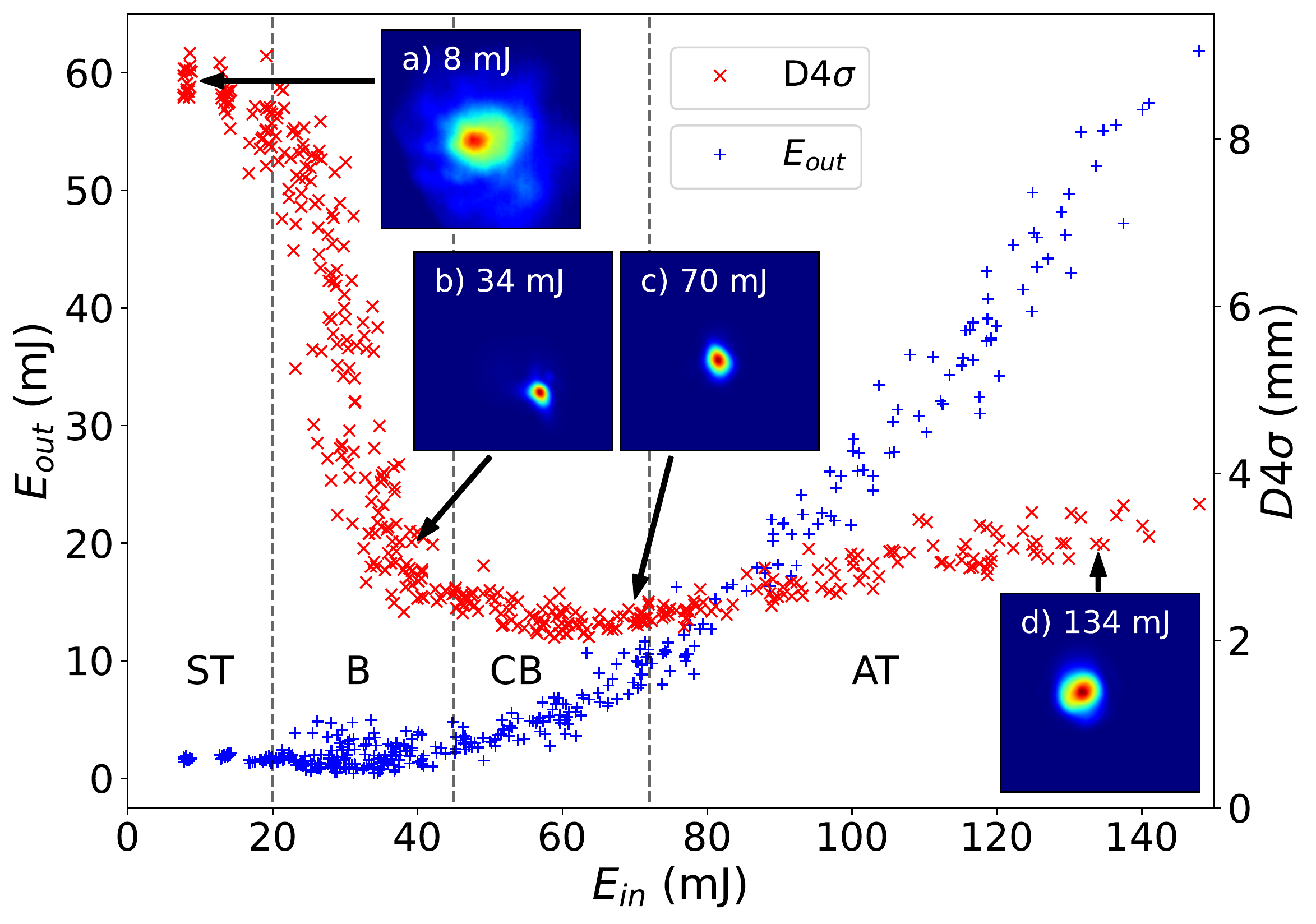}
\caption{Measurement results for transmitted pulse D4$\sigma$ width (right $y$-axis, red crosses) and energy $E_{out}$ (left $y$-axis, blue plus symbols) as a function of pulse input energy for resonant 780 nm pulses and $\mathcal{N}=7\cdot 10^{14}\mathrm{~cm}^{-3}$ vapor density. Dashed vertical lines mark approximate domain boundaries, 
insets show representative transmitted pulse shapes: a) pulse in the sub-threshold (ST) domain, b) narrow-width pulses in the breakthrough (B) and c) confined beam (CB) domains and finally d) widening pulse of the asymptotic transparency (AT) domain.}
\label{fig_samples7res}       
\end{figure}

As a representative example of the measurement results for the transmitted pulse properties we obtained, 
Fig. \ref{fig_samples7res} shows the width of the transmitted pulse together with 
its pulse $E_{out}$ as a function of $E_{in}$, measured for $\mathcal{N}=7\cdot 10^{14}\mathrm{~cm}^{-3}$ vapor density, resonant ($780$ nm) pulses. 
To characterize the beam width, we use the  D4$\sigma$ width (i.e. the second moment width) of the fluence profile $\mathcal{F}(x,y)$ defined as:
\begin{equation}
 D4\sigma=4\sqrt{
 \frac{\int\left(\mathcal{F}(x,y)(x-\bar{x})^2+\mathcal{F}(x,y)(y-\bar{y})^2\right)dxdy}
 {\int\mathcal{F}(x,y)dxdy}}
\end{equation}
where
\begin{equation}
 \bar{x}=\frac{\int x\mathcal{F}(x,y)dxdy} {\int\mathcal{F}(x,y)dxdy} \mathrm{~and~}
 \bar{y}=\frac{\int y\mathcal{F}(x,y)dxdy} {\int\mathcal{F}(x,y)dxdy}.
\end{equation}
$x$ and $y$ are the coordinates in the camera plane in these formulas, 
$\bar{x},\bar{y}$ are coordinates of the geometric center and 
clearly $\int\mathcal{F}(x,y)dxdy = E_{out}$.
The figure depicts ``raw'' measurement data, each marker corresponds to a single measurement. 
Insets show transmitted pulse camera pictures for a few single representative measurements. 
Several regimes are visible on the plots, as discussed in \cite{Demeter2021}.
For small $E_{in}$ in the \textit{sub-threshold} domain (marked ``ST'' on Fig. \ref{fig_samples7res}), $E_{out}$ is below the noise-floor of the
measurement and the image on the pickoff camera is very broad (see inset a) ). Then follows the sharp \textit{breakthrough transition} (marked ``B'' on Fig. \ref{fig_samples7res}), where transmitted pulse width drops quickly as it develops a sharp, narrow, high-fluence feature and $E_{out}$ starts to increase. Just above this is the \textit{confined beam} domain (``CB'' on Fig. \ref{fig_samples7res}), where the pulse width slowly approaches a minimum, while $E_{out}$ increases (insets b) and c) ). In the final region named the
\textit{asymptotic transparency} domain (``AT'' on Fig. \ref{fig_samples7res}), both $E_{out}$ and pulse width increase steadily (inset d) ). This region
is associated with the saturation of the optical nonlinearity of the medium due to complete one-electron ionization in the plasma column core ($P_{max}=1$).

\begin{figure}[htbp]
\centering
\includegraphics[width=1\columnwidth]{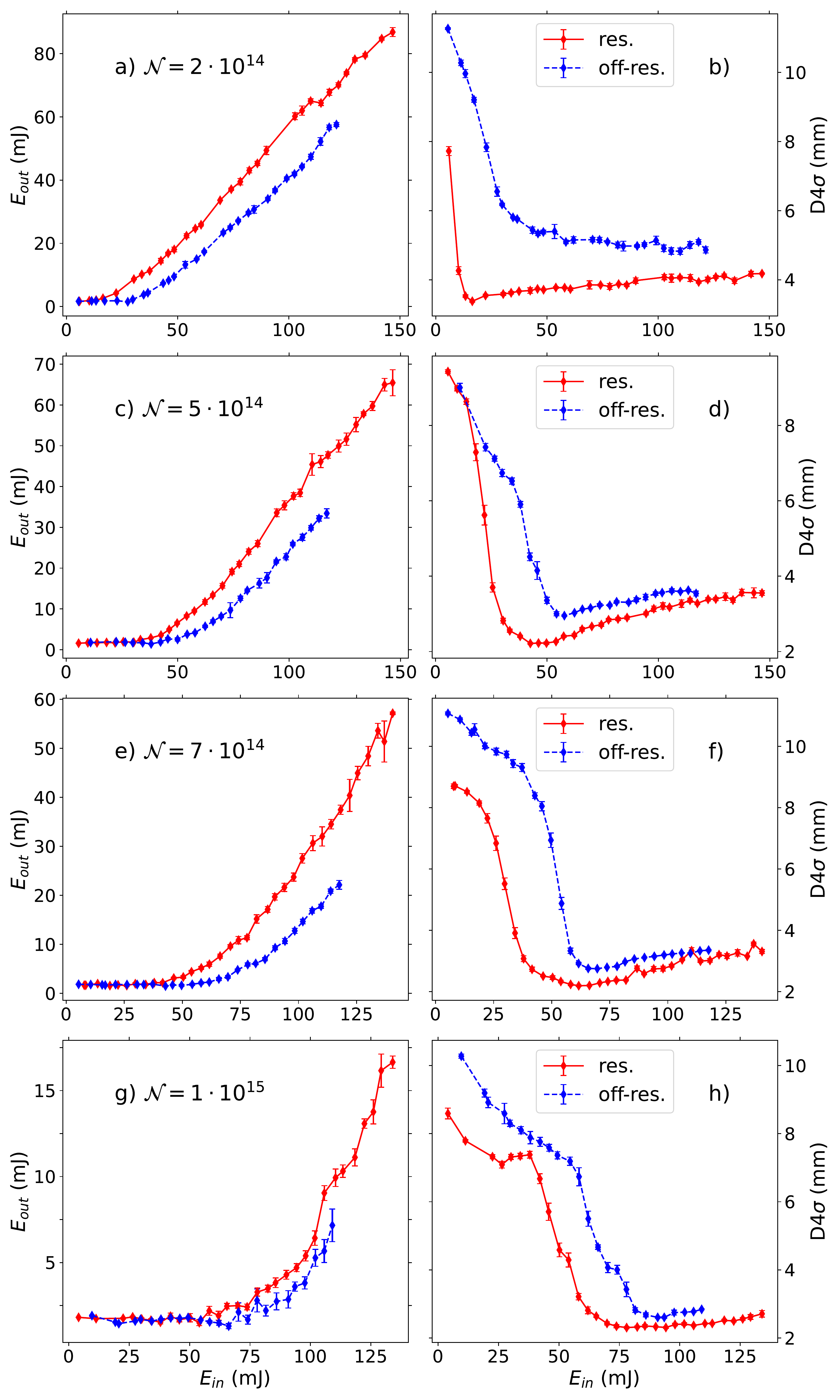}
\caption{Transmitted pulse energy $E_{out}$ (left panels) and D4$\sigma$ width (right panels) as a function
of pulse energy for all four vapor densities and both ionizing pulse wavelengths. Markers show binned data averages, error bars depict the standard error of the mean. Red symbols/lines mark 780 nm resonant, blue lines/symbols mark 810 nm off-resonant measurements.  a), b)  $\mathcal{N}=2\cdot 10^{14}\mathrm{~cm}^{-3}$;
c), d)  $\mathcal{N}=5\cdot 10^{14}\mathrm{~cm}^{-3}$;
e), f)  $\mathcal{N}=7\cdot 10^{14}\mathrm{~cm}^{-3}$;
g), h)  $\mathcal{N}=1\cdot 10^{15}\mathrm{~cm}^{-3}$.
}
\label{fig_D4sigma}       
\end{figure}

The main results of the propagation experiments can be seen on Figs. \ref{fig_D4sigma} where $E_{out}$ and  transmitted pulse D4$\sigma$ width can be seen for all four vapor densities studied and both ionizing pulse central wavelengths. Binned data averages are plotted with error bars marking the standard error of the mean. The figures show that off-resonant ionizing pulses (blue lines) behave similarly to  resonant pulses (red lines) in general. However, $E_{out}$ is smaller for any given $E_{in}$. The
breakthrough transition requires higher $E_{in}$ and the transmitted pulse is always wider for off-resonant pulses.
As the vapor density increases, the breakthrough transition shifts to higher $E_{in}$ for both wavelengths.
Apart from the lowest density measurements, the off-resonant pulse also acquires a minimum width after the breakthrough. The larger $E_{out}$ and smaller D4$\sigma$ width combine to give rise to a substantially larger peak fluence in the resonant case than in the off-resonant case. Overall, Figs. \ref{fig_D4sigma} b), d), f) and h) show that the nonlinear self-focusing effect of the vapor is stronger on the resonant pulses than on the off-resonant pulses, in accordance with theoretical predictions.

\begin{figure}[htb]
\centering
\includegraphics[width=0.85\columnwidth]{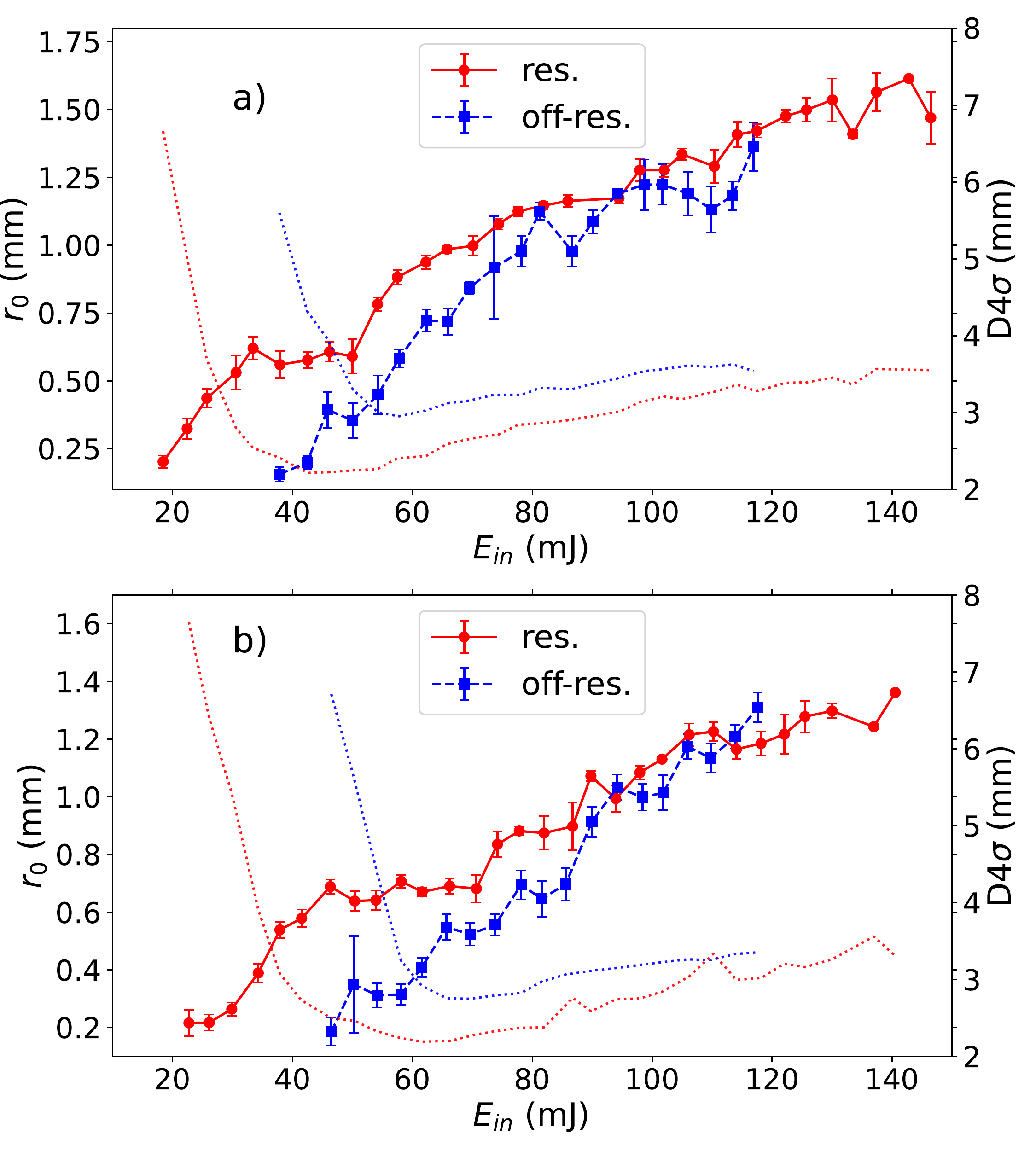}
\caption{Plasma core radius $r_0$ for a) $\mathcal{N}=5\cdot 10^{14}\mathrm{~cm^{-3}}$ vapor density and b) $\mathcal{N}=7\cdot 10^{14}\mathrm{~cm^{-3}}$ vapor density (left vertical axis). Solid red line / symbols mark resonant, blue dashed line / symbols mark off-resonant measurements. Transmitted pulse D4$\sigma$ width is also shown for reference with dotted lines (right vertical axis, red / blue for resonant / off-resonant). }
\label{fig_plasmaradius}       
\end{figure}

\begin{figure}[htb]
\centering
\includegraphics[width=1\columnwidth]{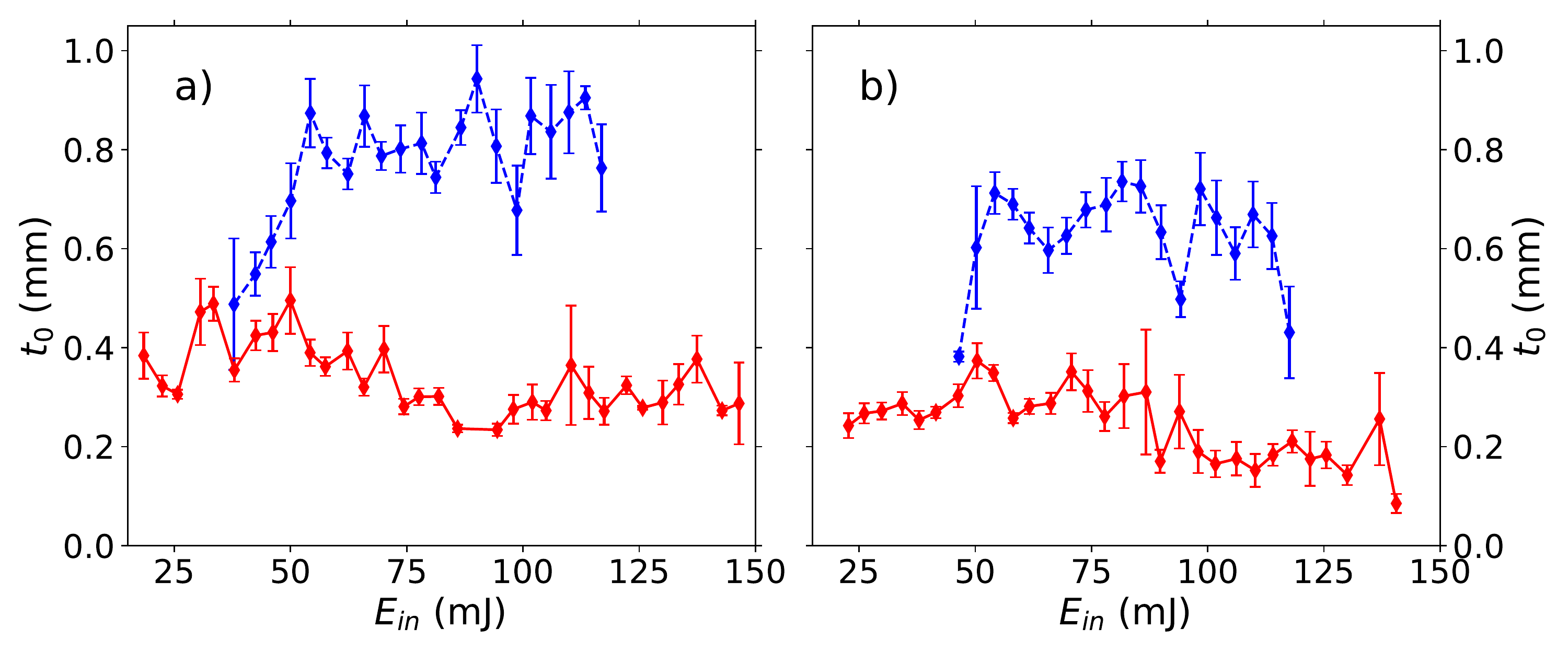}
\caption{Plasma sheath width $t_0$ for a) $\mathcal{N}=5\cdot 10^{14}\mathrm{~cm^{-3}}$ vapor density and 
b) $\mathcal{N}=7\cdot 10^{14}\mathrm{~cm^{-3}}$. Red solid lines correspond to resonant, dashed blue lines correspond to off-resonant pulses.}
\label{fig_plasmachannel}       
\end{figure}

We used schlieren imaging to measure plasma column dimensions for the two intermediate vapor densities together with the propagation measurement.
Figure \ref{fig_plasmaradius} shows the plasma core radius $r_0$ we obtained, as a function of $E_{in}$. Transmitted pulse D4$\sigma$ width is also shown on the figure for reference.
It is visible that
a measureable $r_0$ appears at the breakthrough transition where the transmitted pulse width drops drastically. For resonant pulses, following an initial sharp increase of $r_0$ there is a visible ``shoulder'' of near constant plasma radius, approximately corresponding to the confined beam domain ($E_{in}=32-50 \mathrm{~mJ}$ and $E_{in}=45-72 \mathrm{~mJ}$ for the two vapor densities depicted). There
is no such feature visible for off-resonant pulses. The plasma column reaches the downstream end
of the 10 meter vapor source at higher $E_{in}$ for off-resonant pulses (as the breakthrough requires higher energies),
there is a roughly $r_0=0.5-0.6$ mm plasma radius with resonant pulses already
when the first detectable plasma appears for the off-resonance case. However,
by the end of the input energy range both resonant and off-resonant pulses produce a plasma core of roughly equal radius.
The most important difference between the two cases is the plasma sheath width $t_0$, which is much smaller for resonant pulses (Fig. \ref{fig_plasmachannel}). Though this quantity fluctuates a lot more than $r_0$, especially for the off-resonant case, it is clear that it is at least 2-3 times as large for off-resonant pulses.

\begin{figure}[htb]
\centering
\includegraphics[width=1\columnwidth]{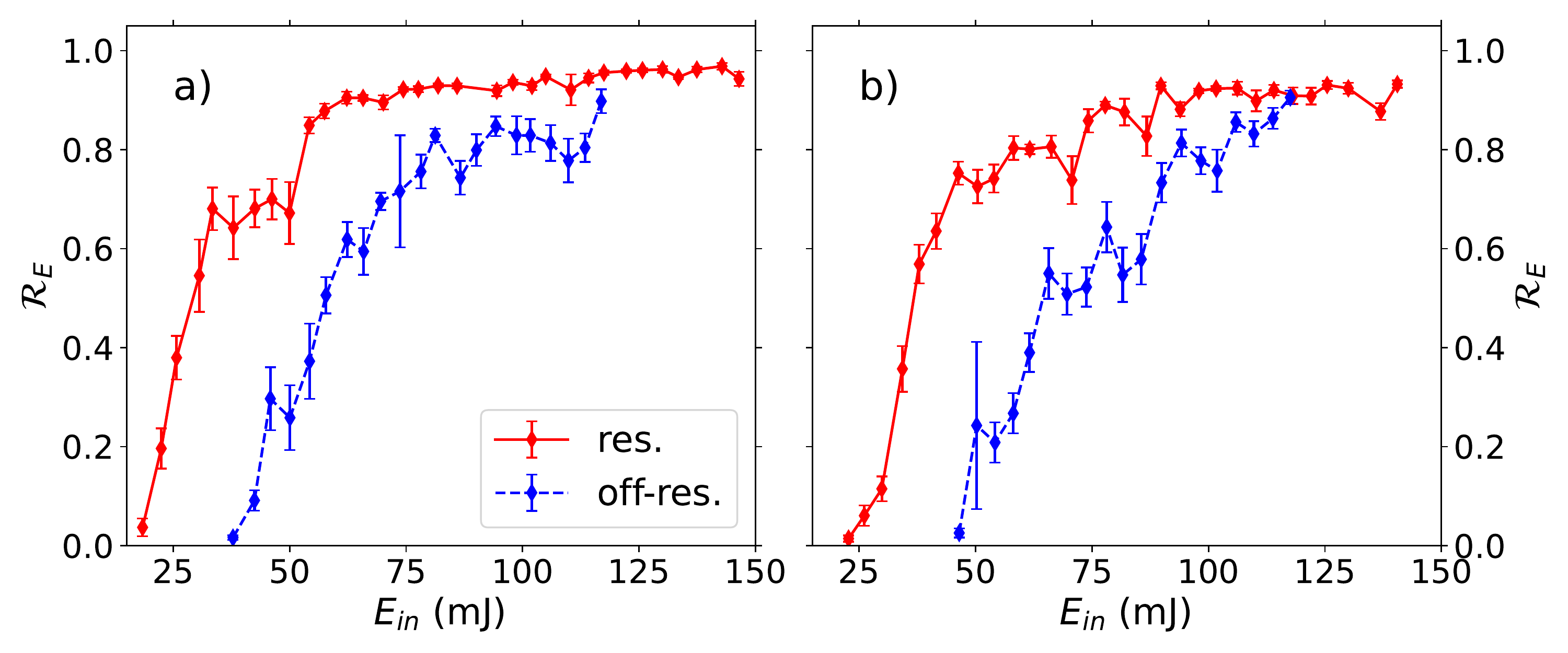}
\caption{Pulse energy ratio $\mathcal{R}_e$ within the $r_0$ radius plasma column core for a) $\mathcal{N}=5\cdot 10^{14}\mathrm{~cm^{-3}}$ vapor density and 
b) $\mathcal{N}=7\cdot 10^{14}\mathrm{~cm^{-3}}$. Red solid lines correspond to resonant, dashed blue lines correspond to off-resonant pulses.}
\label{fig_plasma_Eratio}       
\end{figure}

 Figure \ref{fig_plasmaradius} shows that 
plasma is detected for some shots already at the start of the breakthrough domain. These shots are rare initially, but their frequency increases as $E_{in}$ progresses towards the confined beam domain. The D4$\sigma$ width corresponding to the shots with plasma is small, whereas the shots where plasma is not detected, are much wider. As $E_{in}$ increases, shots with narrow D4$\sigma$ width and measurable plasma become more frequent and the average transmitted pulse width drops quickly. By the end of the breakthrough domain, all shots produce a measurable plasma column at the end of the vapor source.
Note that because our evaluation of the schlieren images yields no direct information on the ionization fraction $P_{max}$, we only assume that we have $P_{max}\approx 1$ at the plasma core if $r_0$ is substantial (i.e. we have a plateau of the ionization fraction in the center). Naturally, 'substantial' must be in comparison with $t_0$, i.e we may safely assume $P_{max}= 1$, only if $r_0\gtrsim t_0$. This regime is reached much sooner with resonant pulses (also) because of the smaller $t_0$. In fact for resonant pulses we definitely reach it by the start of the confined beam domain.

To estimate the significance of a thinner sheath, we may consider the 
ionization probability profile $P_{ion}(r)$ described by Eqn. \ref{eq_plasmadensity} integrated in a plane perpendicular to the propagation direction. The energy loss of the propagating pulse due to ionization per unit distance $dE^{ion}_{loss}/dz$ will be proportional to this quantity:
\begin{equation}
 \int P_{ion}(r)rdrd\phi=r_0^2\pi+t_0^2\pi+\pi\sqrt{\pi}r_0t_0 \sim \frac{dE^{ion}_{loss}}{dz}
 \label{eq_Pintegral}
\end{equation}
Here we assume that every ionization event removes exactly three photons from the field of the laser pulse --- the smallest number required for ionization at these wavelengths. (The effects of above-threshold ionization with four or more photons are thus excluded from this simple consideration, as is the energy loss by other means e.g. atoms not ionized but left in an excited electronic state.)  Using values for $r_0$ and $t_0$ from Figs. \ref{fig_plasmaradius} and \ref{fig_plasmachannel} that correspond to $E_{in}=100$ mJ and inserting in Eq. \ref{eq_Pintegral}, we can readily see that $dE^{ion}_{loss}/dz$ is roughly twice as large for the off-resonant pulse at this point. For lower (higher) pulse energies, where the ratio $t_0/r_0$ is higher (lower), the ratio by which  
$dE^{ion}_{loss}/dz$ is higher for off-resonant pulses will be higher (lower).

An interesting question regarding pulse propagation is the actual ratio of the pulse energy that propagates inside the plasma column core. In traditional filamentation in atmospheric gases, most of the energy propagates in the low intensity wings of the propagating pulses around the plasma as the medium is transparent for low-intensity light. In the resonant pulse scenario discussed here, however, there is  absorption for an arbitrarily low intensity and theory predicts that most of the energy is channeled inside the core, where resonant absorption is stopped by the removal of the valence electron. The actual ratio $\mathcal{R}_E$ of pulse energy propagating within the plasma column core can be estimated from the camera images by integrating the fluence distribution within an $r_0$ radius around the center. Figure   
\ref{fig_plasma_Eratio} shows the results of this evaluation. For the resonant case, around 90 \% of the pulse energy is found to be channeled in the core in the asymptotic transparency domain. Off-resonant pulses are channeled less efficiently, especially for low energies. Nevertheless, even they show a behavior similar to resonant pulses at high energy, i.e. they resemble the propagation of resonant pulses much more than they do traditional non-resonant filamentation phenomena.

The experimental data thus proves, that resonant self focusing helps contain laser pulse energy near the plasma column core very effectively. Pulse energy is channeled in the core where the optical nonlinearity and absorption are saturated due to complete single electron ionization. As a consequence, the plasma sheath layer with partial ionization is much thinner for resonant pulses. Because less energy is lost by resonant pulses for the partial ionization, these pulses have a greater penetration depth for a given energy, i.e. they require less energy to create a continuous plasma column of given length.
This is clearly  very advantageous for creating a long plasma column and particularly so for accelerator applications.

\subsection{Considerations for longer plasma columns}

\begin{figure}[htb]
\centering
\includegraphics[width=1\columnwidth]{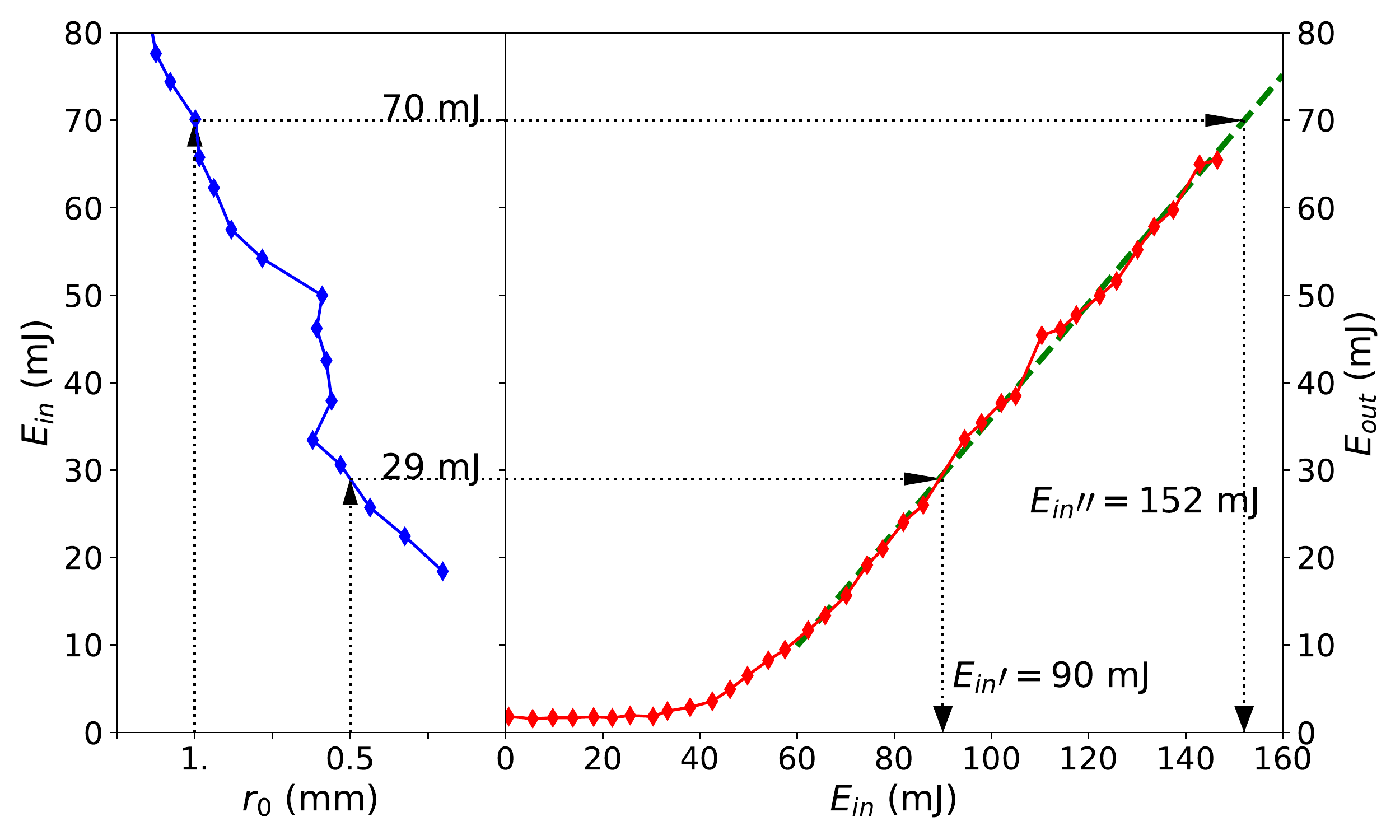}
\caption{Illustration of the recursive method for estimating the energy necessary to create a 20 m long plasma column of a given radius in $\mathcal{N}=5\cdot 10^{14} \mathrm{~cm}^{-3}$ density vapor. Left panel: $r_0(E_{in})$ curve for resonant pulses from Fig. \ref{fig_plasmaradius} a), rotated anticlockwise by 90 degrees. Right panel: $E_{out}(E_{in})$ curve for resonant pulses from Fig. \ref{fig_D4sigma} c). Dashed line in the right panel marks the line of best fit for the upper range of the experimental data.}
\label{fig_recursive1}       
\end{figure}

Using the data we may also estimate the pulse energy that could be capable of 
creating a $z_{max}=20$ m long plasma column. The estimate is based on the observation made in numerical modeling of the system that at a given vapor density and beam focusing, the properties of the propagating pulse (and thus the properties of the plasma column) at a given spatial position will depend only on the energy still remaining in the pulse at that point \cite{Demeter2019,Demeter2021}.
It is made by recursion using the information on Figs. \ref{fig_D4sigma} and \ref{fig_plasmaradius} -- the curves required for the process in this example are shown together in Fig. \ref{fig_recursive1} for convenience. 
If we require a $r_0=0.5$ mm plasma column after $z_{max}=20$ m 
in $\mathcal{N}=5\cdot 10^{14} \mathrm{~cm}^{-3}$ density vapor and consider
resonant ionizing pulses, we first establish that we need $E_{in}=29$ mJ 
to have $r_0=0.5$ mm after $z=10$ m (left panel of Fig. \ref{fig_recursive1}). We now estimate that if we have a pulse which still contains 29 mJ energy after $z=10$ m, that pulse will be able to create the required plasma column in the \textit{second} 10 meter section of the 20-meter-long vapor as well. Finding $E_{in}'$ that yields $E_{out}=29$ mJ from the $E_{out}(E_{in})$ curve we obtain $E_{in}'=90$ mJ (right panel of Fig. \ref{fig_recursive1}, arrows with dotted line depict the process).    
Thus, we estimate that a resonant pulse with $E_{in}\gtrapprox90$ mJ would be able to create the required plasma column in a 20 m long vapor source --- a value well within the capabilities of the laser system.  

The same process can be used to estimate the energy required for a $z_{max}=20$ m long, $r_0=1$ mm plasma column at the same vapor density. Figure \ref{fig_recursive1} left panel shows that
$E_{in}=70$ mJ is required for $r_0=1$ mm after $z=10$ m.  Projecting this value as $E_{out}$ onto the right panel we see that we are somewhat above the range of available experimental data. However, using linear extrapolation, 
we can estimate that $E_{in}''\approx152$ mJ yields $E_{out}=70$ mJ after 10 meters of propagation, which in turn is the energy requirement for a pulse to create a 20 m long, 1 mm radius plasma column. 

\begin{figure}[htb]
\centering
\includegraphics[width=1\columnwidth]{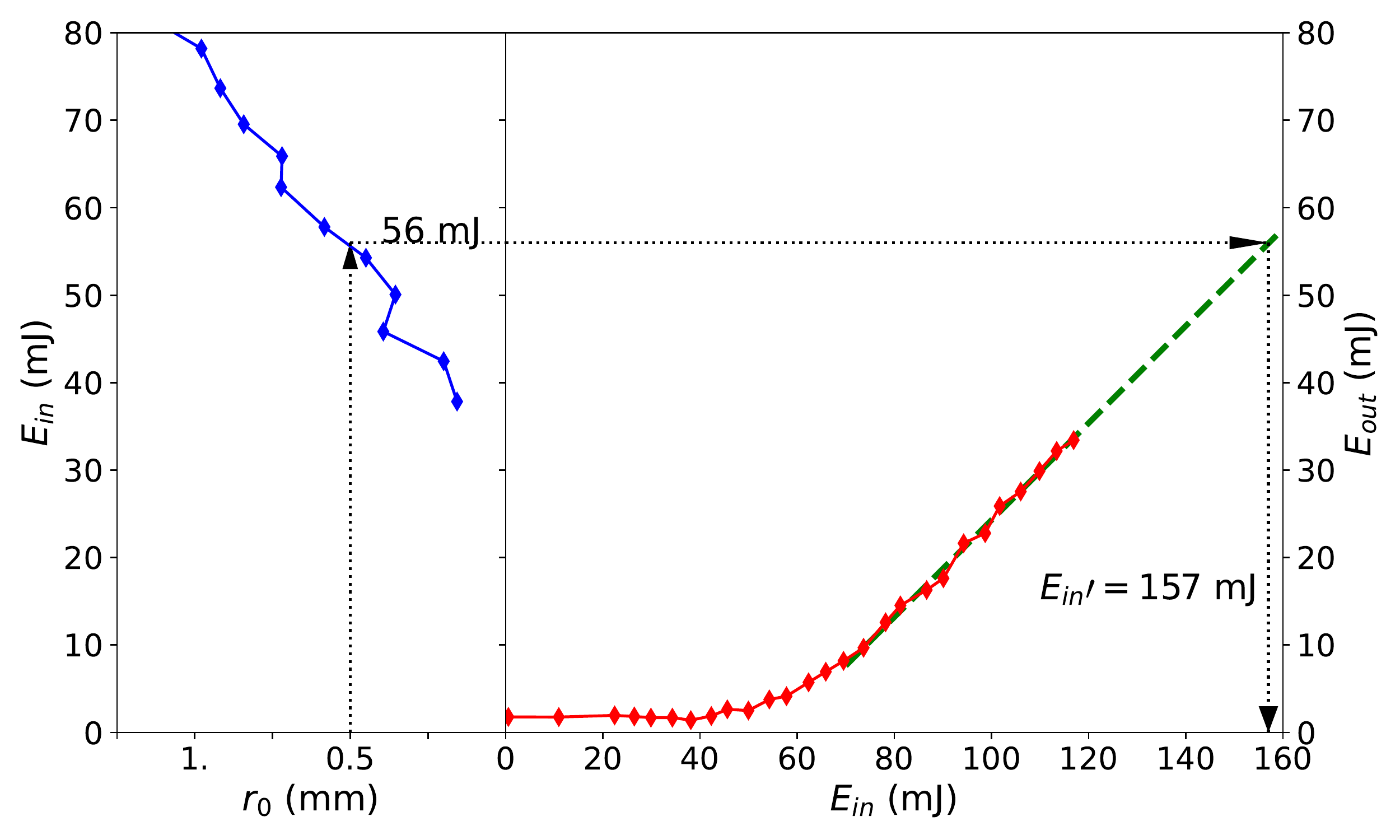}
\caption{Illustration of the recursive method for estimating the energy necessary to create a 20 m long plasma column of a given radius in $\mathcal{N}=5\cdot 10^{14} \mathrm{~cm}^{-3}$ density vapor for off-resonant pulses. Left panel: $r_0(E_{in})$ curve for off-resonant pulses from Fig. \ref{fig_plasmaradius} b), rotated anticlockwise by 90 degrees. Right panel: $E_{out}(E_{in})$ curve for off-resonant pulses from Fig. \ref{fig_D4sigma} c). Dashed line in the right panel marks the line of best fit for the upper range of the experimental data.}
\label{fig_recursive2}       
\end{figure}

Repeating the process to investigate the possibility of generating a $z_{max}=20$ m, $r_0\geq0.5$ mm plasma column by off-resonant pulses, we find that $E_{out}=56$ mJ is required after $z=10$ m propagation (Fig. \ref{fig_recursive2}). However, projecting this value to find $E_{in}'$ we are far above the experimental curve due to the fact that the plasma column appears for larger $E_{in}$ and transmitted energies are smaller for the off-resonant case. Linear extrapolation again yields an estimate for the required
pulse energy to be $E_{in}'\approx157$ mJ, much greater than the 90 mJ value of the resonant case. Also, this estimate is to be handled with greater caution due to its distance from the measured range. The same process with the data for $\mathcal{N}=7\cdot 10^{14} \mathrm{~cm}^{-3}$ density vapor and resonant pulses 
readily yields $E_{in}'\approx118$ mJ for the limit of creating a $z_{max}=20$ m plasma column with $r_0\geq0.5$ mm (without any extrapolation). We also note that given enough experimental data, the recursive process could be repeated to estimate
the pulse energy requirement for 30 meter long or even longer plasma columns.

\section{Simulation}

\subsection{Theoretical framework}

A theory has been developed recently to describe the propagation of ultra-short, ionizing laser pulses through Rb vapor under the condition of single photon resonance from the ground state \cite{Demeter2019}. Results from numerical simulations have been found to agree qualitatively with experimental findings of resonant pulse propagation for the transmitted pulse energy and width \cite{Demeter2021}. 
Using this as a starting point, we have derived a more general theory to treat the propagation of off-resonant pulses and resonant pulses in a unified way and explore the difference between their behavior. Here we present only a very brief outline, as the basic concept is the same as the one described in \cite{Demeter2021} in greater detail.

The basic equation used for the pulse propagating along the $z$ direction is written for the complex envelope function $\mathcal{E}(r,z,t)$ of the axisymmetric laser field 
$E(r,z,t)=\frac{1}{2}\mathcal{E}(r,z,t)\exp(ik_0z-\omega_0t)+c.c.$ ($\mathcal{E}(r,z,t)$
($\omega_0$ and $k_0$ being the central frequency and wavenumber in vacuum).
Standard methods in the treatment of ultrashort pulses are employed \cite{Couairon2011} such as 
the paraxial approximation, transforming to Fourier space with respect to $\tau=t-z/c$, 
the delayed time: $\tilde{\mathcal{E}}(r,z,\omega)=\mathfrak{F}\{\mathcal{E}(r,z,\tau)\}$
and using the Slowly Evolving Wave Approximation (SEWA) \cite{Brabec1997} to arrive at the propagation equation:  
\begin{equation}
\begin{aligned}
\partial_z \tilde{\mathcal{E}}  = &
\frac{i}{2k} \nabla_\perp^2\tilde{\mathcal{E}}
+ i\frac{k}{2\epsilon_0} \tilde{\mathcal{P}} \\
& -\eta_0\hbar\omega_0\mathcal{N}\tilde{\mathcal{Q}}
- \frac{ik}{2}\frac{e^2\mathcal{N}}{\epsilon_0 m_e (\omega_0+\omega)^2} \tilde{\mathcal{R}}
\end{aligned}
\label{waveeq}
\end{equation}
Here $e,m_e$ are the elementary charge and electron mass,
$\epsilon_0$ is the vacuum permittivity and $\eta_0$ the vacuum impedance. 
The first term on the right-handside (RHS) of Eq. \ref{waveeq} is due to diffraction, 
the last term is the plasma dispersion term. 
The third term on the RHS of Eq. \ref{waveeq} is an energy loss term
derived from the requirement that when an atom is ionized, an appropriate number of photons are absorbed from the field (the number depending on the electronic level that the 
valence electron was in prior to ionization).

\begin{figure}[htb]
\centering
\includegraphics[width=0.7\columnwidth]{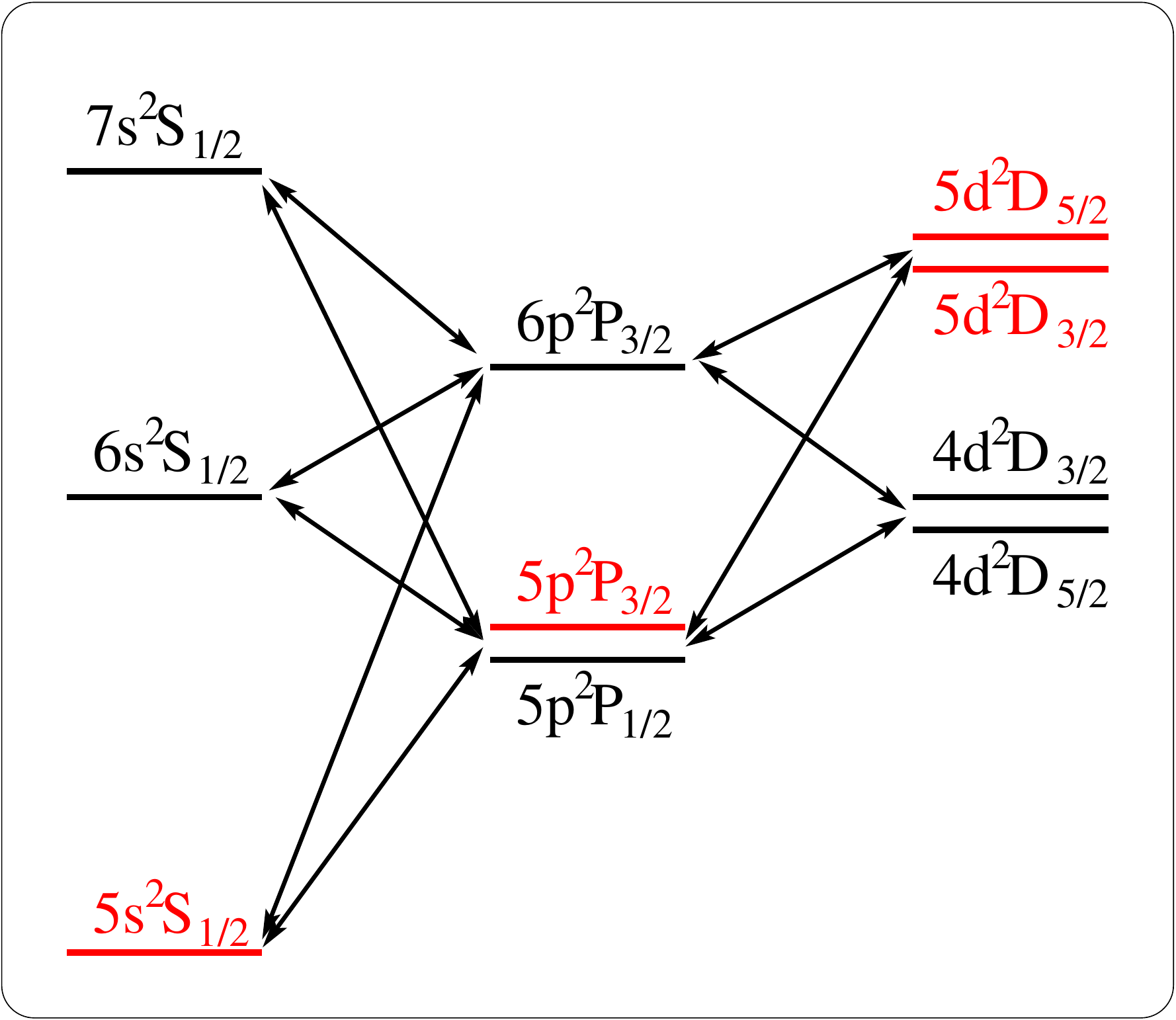}
\caption{a) Electronic levels of the rubidium atom that are included in the theoretical model. The states used in the original, 4-level model \cite{Demeter2019} are highlighted in red. Black arrows mark allowed dipole transitions between the states, but transitions to different fine-structure sublevels are not resolved, i.e. only a single arrow corresponds e.g. to the 780 nm D$_2$ and 795 nm D$_1$ lines from $\mathrm{5s^2S_{1/2}}$ to $\mathrm{5p^2P_{3/2}}$, and $\mathrm{5p^2P_{1/2}}$ states.  }
\label{fig_Rblevels}       
\end{figure}

The second term on the RHS  is the atomic polarization, which, for the resonant case, is dominated by Rabi-oscillation type transitions on single-photon resonances. Contributions 
of this type cannot be expressed in terms of usual optical nonlinear coefficients \cite{BoydNonlinOptics}. Therefore our theory includes an explicit calculation of the atomic state using the Schr{\"o}dinger equation written for the $\alpha_j(t)$ probability amplitudes: 
$|\psi\rangle=\sum_j\alpha_j(t)\exp(-i\omega_jt)|j\rangle$ ($\hbar\omega_j$ is the energy of the energy eigenstate $|j\rangle$). The original model \cite{Demeter2019} included just four atomic states, the ground state and three excited states that are accessible via resonant interaction within the bandwidth of the 780 nm ionizing pulses (see Figs. \ref{fig_spectrum} and \ref{fig_Rblevels}). However, the dominance of these over other atomic states will be much less significant for off-resonant pulses centered around 810 nm. Therefore the model has been expanded to include 10 atomic levels in all (Fig. \ref{fig_Rblevels}). The levels were chosen by first considering a more general model with 18 atomic states and then selecting only those which proved to acquire a maximum occupation probability of at least 0.01 during the interaction with the laser.   
The evolution of the atomic state is thus described by:
\begin{equation}
\begin{aligned}
 \partial_t\alpha_j = & \frac{i}{2\hbar}\sum_k \mathcal{E}e^{-i\Delta_{jk}t} d_{jk}\alpha_k \\
 & + \frac{i}{2\hbar}\sum_{k'} \mathcal{E}^*e^{i\Delta_{jk'}t} d_{jk'}\alpha_{k'}-\frac{\Gamma_j}{2}\alpha_j
 \end{aligned}
 \label{schrodinger}
\end{equation}
The summation for index $k$ runs over the lower lying atomic states ($\hbar\omega_k < \hbar\omega_j$) for which the dipole matrix element
$d_{jk}\neq0$, the summation for $k'$ for higher lying states in a similar manner.
 $\Delta_{jk} = \omega_0 - (\omega_j-\omega_k)$ is the detuning of the laser central frequency from the $|k\rangle\rightarrow |j\rangle$ transition frequency. Material parameters needed in the calculation (level energies and dipole oscillator strengths) have been obtained from the literature \cite{SteckRb85,Safronova2004,NIST}. $\Gamma_j$ are level loss rates due to ionization (intensity dependent), obtained from the so-called PPT-formulas \cite{Perelomov1966, Perelomov1967, Perelomov1967b} and experimental data \cite{Duncan2001}. With the atomic state evolution calculated, the atomic polarization term for Eq. \ref{waveeq} is given by the dipole operator expectation value $\langle\hat{d}\rangle$:
 \begin{equation}
\tilde{\mathcal{P}}= \mathfrak{F}\left\{\mathcal{N}\sum_{kl}\alpha_k^*\alpha_l d_{kl}\right\}.
\end{equation}
where $\mathfrak{F}\{\ldotp\}$ marks the time Fourier transform.

\subsection{Numerical results and discussion}

We performed calculations for a pulse propagation scenario similar to the experiments, with
both resonant and off-resonant pulses. The computations were done for a range of input pulse energies with vapor density $\mathcal{N}=7\cdot 10^{14}\mathrm{~cm^{-3}}$. The input laser pulse was a Gaussian beam, with focal parameters derived from the measured virtual laser line camera fluence distributions as in \cite{Demeter2021}. A hyperbolic secant temporal dependence of the initial field was assumed --- note that this ideal pulse shape leads to a significantly narrower spectral width for the pulse than that actually measured (Fig. \ref{fig_spectrum}).  
Transmitted pulse properties and ionization profiles were plotted after 10 meters of propagation for comparison with experimental results.

\begin{figure}[htb]
\centering
\includegraphics[width=1\columnwidth]{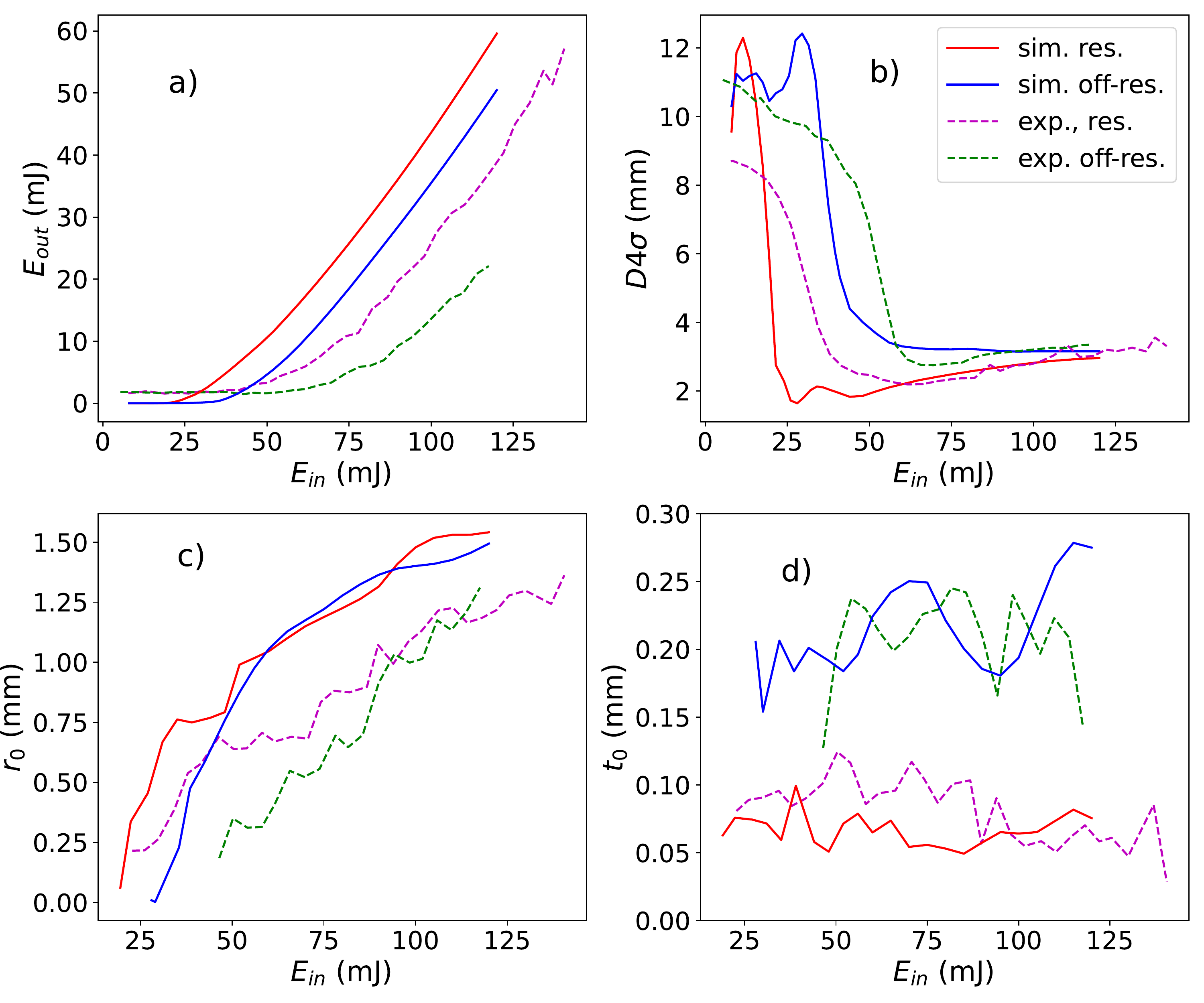}
\caption{Simulation results for $\mathcal{N}=7\cdot 10^{14}\mathrm{~cm^{-3}}$ vapor density a) transmitted pulse energy, b) transmitted pulse $D4\sigma$ width, c) plasma core radius $r_0$ and d) plasma sheath width $t_0$. Solid red lines mark 780 nm resonant results, solid blue lines mark 810 nm off-resonant results. Experimental results are also plotted with dashed lines (without error bars) for an easy comparison. Experimental $t_0$ values on d) are rescaled by a factor of 1/3 to fit close to the range of the simulation results.}
\label{fig_simulation}       
\end{figure}

Figure \ref{fig_simulation} a) and b) show transmitted pulse energies and widths, while 
Fig. \ref{fig_simulation} c) and d) shows plasma core radii and sheath widths. 
There is a good qualitative similarity between experiment and theory albeit with some quantitative discrepancy. Transmitted pulse energy (Fig. \ref{fig_simulation} a) ) is predicted to be significantly larger for any given $E_{in}$ than that measured, but the fact that the resonant pulse $E_{out}$ curve commences growing earlier and is greater than its off-resonant counterpart is clearly represented in the results. Transmitted pulse width (Fig. \ref{fig_simulation} b) ) also shows that simulation reproduces well the breakthrough behavior of the pulses, the quick drop in $D4\sigma$ to a minimum width,  as well as the following expansion of the width of the resonant pulse. The minimum width value obtained in the simulation is somewhat smaller for the resonant case.
The plasma $r_0$ behavior is also well reproduced (Fig. \ref{fig_simulation} c) ), the curve for the resonant case displaying the shoulder of near constant $r_0$ that corresponds approximately to a similar region in the $D4\sigma$ width (the confined beam domain). Finally, the sheath width $t_0$ (Fig. \ref{fig_simulation} d) ) that we obtained from simulation also shows that the resonant pulse gives a $t_0$ value several times smaller than the off-resonant pulse does. However, the numerical values from the simulation are smaller by a factor of about three. (Experimental values have been rescaled on Fig. \ref{fig_simulation} d), see also Fig. \ref{fig_plasmachannel} b).)  

\begin{figure}[htb]
\centering
\includegraphics[width=1\columnwidth]{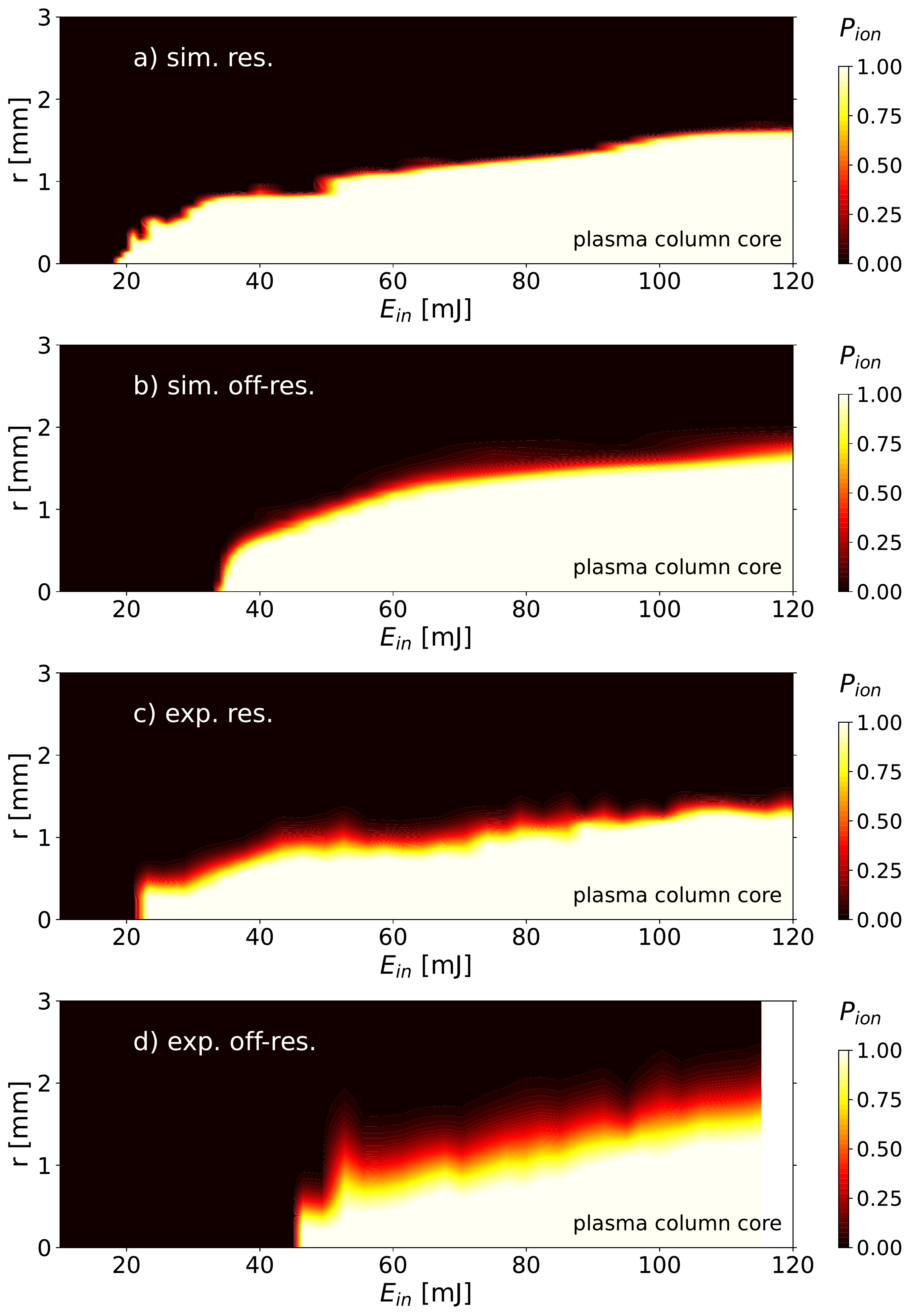}
\caption{Contour plot of the ionization probability at the vapor source exit obtained from a), b) numerical simulation and c), d) schlieren imaging measurements for $\mathcal{N}=7\cdot10^{14}\mathrm{~cm^{-3}}$ density and both resonant and off-resonant pulses.}
\label{fig_sim_plasma_at_exit}       
\end{figure}

It is noteworthy that the drop in transmitted pulse $D4\sigma$ is much sharper for the simulation than measured experimentally. It has been shown in \cite{Demeter2021}, that the parameters of the ionizing pulses (measured beam waist parameters $w_0$, $z_0$ and the beam profile shape as well) fluctuate somewhat shot-to-shot. The simulations on the other hand are performed with constant beam parameters. Varying the beam parameters in the simulation (within the range of variation observed in the experiment) the sharp drop is smeared out and small-scale features (present predominantly for the resonant pulse case) are smoothed, averaged out \cite{Demeter2021}.

\begin{figure}[htb]
\centering
\includegraphics[width=1\columnwidth]{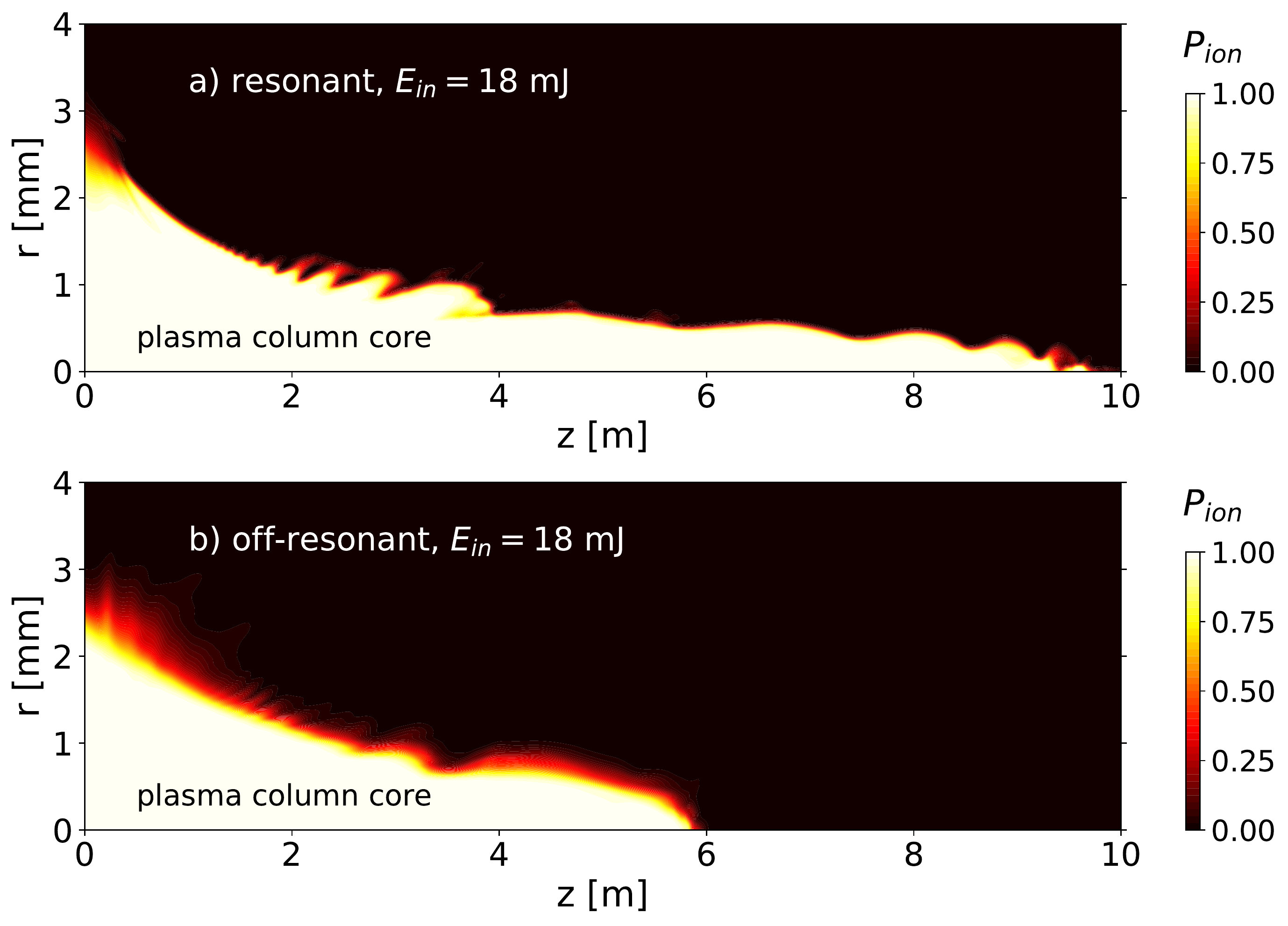}
\caption{Ionization probability inside the vapor source from simulation calculated for a $E_{in}=18$ mJ energy, a) 780 nm resonant and b) 810 nm off-resonant pulse. }
\label{fig_sim_plasmachannel}       
\end{figure}

To visualize the plasma column at the vapor source exit, we show a 2D plot of the ionization probability (i.e. the plasma density) as a function of $E_{in}$ and the radius on Fig. \ref{fig_sim_plasma_at_exit}. Plotted are: a), b) $P_{ion}(E_{in},r)$ at $z=10$ m obtained from the simulation and c), d) ionization probabilities of the form Eq. \ref{eq_plasmadensity}, with parameters $r_0,t_0$ derived from bin averages of binned $r_0$ and $t_0$ values obtained from the experiment (Figs. \ref{fig_plasmaradius} and \ref{fig_plasmachannel}).
The plasma column core in the figures is the region close to the $r=0$ axis where $P_{ion}=1$. The plots show good resemblance between simulation and experiment --- 
the fact that the experimental sheath layer width is much wider than the simulated one is also visible. Finally, 
Figure \ref{fig_sim_plasmachannel} shows two plots for simulated $P_{ion}(z,r)$ calculated for the entire length of the vapor source in case of a resonant and an off-resonant pulse with the same energy.
The pulse energy is low, a little below the breakthrough transition of the resonant pulse. The plots reflect very well that conclusions drawn in the previous sections from quantities observed / calculated for the vapor source exit hold during the entire propagation except for a small transient region just after the entry.  
They depict explicitly that the resonant pulse of equal energy creates a longer plasma column than the off-resonant one.

Finally, some comments on the possible causes of the large quantitative discrepancy between numerically predicted and experimentally observed $E_{out}$ and $t_0$ values.
In several respects, the simulation assumes a situation that is only imperfectly realized in the experiment. First, the real laser pulse is not axially symmetric and its spatial distribution is far from being a pure Gaussian beam. While non-axisymmetric (i.e. full 3D) simulations are far out of scope, similar calculations have been conducted by us previously with flattened Gaussian beams of various orders \cite{Gori1994} using the previous numerical model  for resonant propagation \cite{Demeter2019}.These studies determined that the different transverse spatial dependence did not affect the propagation substantially apart from a fairly short transition region at the beginning of the vapor and did not result in a substantial difference in the transmitted energy. Second, the idealized sech temporal pulse shape does lead to a much narrower spectral width of the pulse than that measured for the experimental input (Fig. \ref{fig_spectrum}). This may possibly cause a more significant difference. Furthermore, the simplified calculation of ionization (required to get an atomic model calculable in 2D propagation simulations) may also cause a significant quantitative difference from the experiment.

\section{Summary}

We have presented both experimental and simulation results for the propagation of resonant and off-resonant, ultra-short, ionizing laser pulses in rubidium atomic vapor. 
The 780 nm central wavelength pulses were resonant with atomic lines of the rubidium atom, while the 810 nm central wavelength off-resonant pulses were just above the resonance wavelengths. We performed measurements varying the input pulse energy from 0 to $\sim$150 mJ and repeated the measurements for several vapor densities.

We measured the transmitted pulse energy and width after propagation along the 10-meter-long rubidium vapor column. Simultaneously, we have investigated the transverse extent of the plasma column created by the pulses close to the downstream end of the vapor column using schlieren imaging. From the schlieren images, we determined the radius of the plasma column core, where the laser pulses achieve one-electron ionization of the rubidium atoms with a probability very close to 1. We also determined the plasma column sheath layer width, which characterizes the fall-off distance of the plasma density from the core to the unionized, neutral vapor.

We established that resonant pulses lose less energy during propagation per unit distance and are confined more strongly in the transverse plane by the interaction with the vapor. 
For low energy, resonant ionizing pulses create a plasma with significantly larger core radius at the vapor's end than off-resonant pulses do. For high energy pulses, the radius of the plasma core is approximately equal for the two cases. For the entire energy range investigated, resonant pulses create a plasma column with a much thinner sheath layer. Based on our results, we conclude that the greater energy loss in case of off-resonant pulses comes predominantly from this wider sheath layer, as they leave a much larger volume of partially ionized atoms around the core. 
For this reason, resonant pulses are able to create a longer plasma column than off-resonant pulses with the same energy. 
We observe that, contrary to traditional filamentation in atmospheric gases, both resonant and off-resonant pulses are channeled in the plasma core where further ionization and thus energy absorption does not take place. However, resonant pulses are being channeled much more efficiently with around 90\% of the pulse energy traveling in the core for mid- to high-energy pulses.

 We have also used our experimental results to estimate the pulse energy required for creating a 20-meter-long plasma column under identical conditions. While the laser used in our experiments would be able to create this double length plasma column with resonant pulses, it is far from it when off-resonant pulses are used.

We generalized a theoretical description developed for the resonant pulse propagation problem to treat the off-resonant case as well. Comparing simulation results with measurement, we conclude that 
they are qualitatively similar, even if the quantitative predictive power of the theory is still lacking in certain respects. Using simulations results, we have shown that the plasma column created by resonant pulses is bounded more sharply almost all along the entire vapor column and that this allows the resonant pulses to generate a much longer plasma column with equal initial pulse energy.   
Our results show that single-photon resonances have a major effect on pulse propagation and that they can be very advantageous if the requirement is to generate a long, 10-meter-scale plasma column. Our results can potentially be very significant for the construction of plasma wakefield accelerator devices, but may also be of interest for other applications with high-power laser pulses, e.g. lightning protection or remote sensing applications.

\section*{Funding}
The research was supported by the Hungarian National Research, Development and Innovation Office (NKFIH) under the contract numbers NKFIH 2019-2.1.6-NEMZ\_KI-2019-00004 and MEC\_R 140947.
On behalf of Project Awakelaser we are grateful for the usage of ELKH Cloud \cite{Heder2022} which helped us achieve the results published in this paper.

\section*{CRediT authorship contribution statements}

{\bf G. Demeter:} Conceptualization; Investigation; Formal analysis;
Methodology; Software; Data Curation; Visualization; Writing - original draft;
{\bf J. T. Moody:} Conceptualization; Methodology; Resources; Supervision;
Investigation; Data Curation; Writing - review \& editing;
{\bf M. Á. Kedves:} Conceptualization; Investigation; Data Curation; Writing -
review \& editing;
{\bf F. Batsch:} Resources; Writing - review \& editing;
{\bf M. Bergamaschi:} Resources; Writing - review \& editing;
{\bf V. Fedosseev:} Resources; Writing - review \& editing;
{\bf E. Granados:} Resources; Writing - review \& editing;
{\bf P. Muggli:} Project Administration; Funding acquisition; Writing - review \& editing;
{\bf H. Panuganti:} Resources; Writing - review \& editing;
{\bf G. Zevi Della Porta:} Resources; Writing - review \& editing;

\section*{Declaration of competing interests}

The authors declare that they have no known competing financial
interests or personal relationships that could have appeared to
influence the work reported in this paper.

\clearpage

\bibliography{/home/gdemeter/fiz/manuscript/bibliography_pulseprop}

\bibliographystyle{apsrev4-1}


\end{document}